\documentclass[journal]{IEEEtran}

\usepackage{amssymb,amsmath,amsfonts,amsthm}
\usepackage{mathrsfs}
\usepackage{cite}
\usepackage{array}
\usepackage{tabularx}
\usepackage{color}
\usepackage{mathtools}
\usepackage{subfigure}
\usepackage{graphicx}
\usepackage{bbm}
\usepackage{dsfont}
\usepackage{setspace}
\usepackage{multirow}
\DeclareMathOperator{\sinc}{sinc}

\newtheorem{definition}{Definition}

\newtheorem{propi}{Proposition}

\newtheorem{remark}{Remark}
\newcommand{\argmax}{\mathop{\rm arg~max}\limits}

\IEEEoverridecommandlockouts

\begin{document}
\title{Decentralized Opportunistic Access for D2D Underlaid Cellular Networks}
\author{Zheng~Chen and~Marios~Kountouris
\thanks{This work was presented in part in 2014 IEEE Global Communications Conference \cite{globecom14} and in 2015 IEEE International Conference on Communication Workshop \cite{two_tier}.}
\thanks{Z. Chen is with the Laboratoire de Signaux et Syst\`{e}mes (L2S, UMR8506),
CentraleSup\'{e}lec - CNRS - Universit\'{e} Paris-Sud,
Gif-sur-Yvette, France. Email: zheng.chen@centralesupelec.fr.}
\thanks{M. Kountouris is with the Mathematical and Algorithmic Sciences Lab, France Research Center, Huawei Technologies Co. Ltd. Email: marios.kountouris@huawei.com.} 
}

\maketitle
\begin{abstract}
We propose a decentralized access control scheme for interference management in D2D (device-to-device) underlaid cellular networks. Our method combines SIR-aware link activation with cellular exclusion regions in a case where D2D links opportunistically access the licensed cellular spectrum. Analytical expressions and tight approximations for the coverage probabilities of cellular and D2D links are derived. We characterize the impact of the guard zone radius and the SIR threshold on the D2D area spectral efficiency and cellular coverage. A tractable approach was proposed in order to find the SIR threshold and guard zone radius, which maximize the area spectral efficiency of the D2D communication while ensuring sufficient coverage probability for cellular uplink users. Simulations validate the accuracy of our analytical results and show the performance gain of our proposed scheme compared to existing state-of-the-art solutions.
\end{abstract}


\section{Introduction}
Current cellular networks are facing immense challenges to cope with the ever increasing demand for throughput and coverage owing to the growing amount of mobile traffic. Traditional methods for boosting the network capacity are either cost ineffective or with limited potential gains and scalability virtues. Device-centric architectures and smarter devices have been identified as disruptive technologies directions, which will lead to fundamental changes in the design of future fifth generation (5G) cellular networks \cite{Boccardi5G}. Leveraging the physical proximity, device-to-device (D2D) communication has the potential to handle local communication more efficiently and the ability to offload cellular traffic. The integration of D2D communications in cellular networks has been proposed as a promising method for increasing spatial resource reuse and effectively support content delivery without significantly harming cellular traffic. 
D2D communication may also enhance link coverage, provide higher area spectral efficiency, and reduce end-to-end latency, while enabling new location-based services and reliable public safety communications. An overview of D2D proximity services in 3GPP standardization activities and of the main challenges in designing D2D-enhanced cellular standards is given in \cite{overview}. 

Despite the potential benefits, there are still significant challenges to their successful implementation, including interference management, self-organization capabilities, network discovery, and distributed resource allocation \cite{survey}.
Direct communication between two devices without traversing the base station (BS) or core network can occur on licensed cellular spectrum (\emph{inband}) or unlicensed spectrum (\emph{outband}).
Most existing work on D2D communications focus on \emph{inband} D2D and especially on D2D underlaid cellular networks where D2D users opportunistically access the licensed spectrum utilized by cellular users~\cite{d2d_resource, mode_selection_power, spectrum_sharing}. In these network deployments, cellular links experience cross-tier interference from co-channel D2D transmissions, whereas D2D pairs experience both inter-D2D interference and cross-tier interference from cellular transmissions.

\subsection{Related Work}
There has been considerable interest in interference management techniques for D2D underlaid cellular networks. Power control and opportunistic medium access control are two effective approaches to harness interference in dense D2D networks. Several proposed techniques have been inspired by threshold scheduling \cite{Weber07}, spatial opportunistic ALOHA \cite{opportunistic_aloha} and opportunistic channel probing \cite{ opportunistic_scheduling}, which have been studied in the context of wireless ad hoc networks.
Several D2D power control strategies are developed and evaluated using the deterministic network deployment model for optimizing different performance metrics \cite{vtc, spectrum_optimization, dynamic, janis2009device, overlaid, operator_device, rate_splitting}.  
Spectrum sharing between ad hoc and cellular networks using a random network model was studied in \cite{jsac_huang}, while power control in wireless ad hoc networks has been studied in \cite{power_femto, fractional, zhang2012random}. In \cite{power_control}, a random network model for D2D underlaid cellular networks is proposed and channel-aware power control and link activation algorithms are developed. 
Different opportunistic access control and link activation techniques for two-tier femtocell networks are proposed in \cite{wcnc_opportunistic}. Therein, contrary to \cite{power_control},  signal-to-interference ratio (SIR) knowledge is exploited to further increase the aggregate throughput. Interferer-channel aware scheduling for large-scale ad hoc networks is investigated in \cite{interference_channel} and in \cite{flashlinq} a synchronous distributed scheduler for peer-to-peer ad hoc networks is proposed.  
None of these existing works on threshold-based scheduling takes into account the aggregate interference that a potential link receives from all concurrent transmissions. Moreover, performance analysis and optimization of SIR-aware opportunistic access has not considered.

Guard zones (exclusion regions) around cellular receivers have been considered in D2D underlaid cellular networks as a means to boost the D2D throughput without significantly degrading the quality of service (QoS) of the cellular network \cite{guard_zone, geometry_based, dynamic_tdd, exclusion_d2d, two_tier}. 
Guard zone based protocols lead to inter-tier dependence among transmitters, i.e. BSs belonging to different tiers exhibit repulsion. 
Recent results on \textit{Poisson Hole Process} \cite{poisson_hole} can be used to calculate interference and coverage probability in guard-zone based two-tier networks.
A cognitive radio network model where no secondary users can lie within the guard zones of primary users is considered in \cite{cognitive}, where bounds on the outage probability are provided. Similar analysis for heterogeneous cellular networks with intra-tier and inter-tier dependence can also be found in \cite{dependence}. Nevertheless, none of these works have considered decentralized D2D access and link activation combined with cellular exclusion regions so as to maximize the D2D throughput subject to cellular coverage constraints.

\subsection{Contributions}
In this paper we consider a multi-cell D2D underlaid cellular network, in which uplink cellular users intend to communicate with their nearest BS while multiple D2D links coexist in the same spectrum. The locations of cellular BSs and potential D2D transmitters are both modeled using a spatial homogeneous Poisson point process (PPP).
A random spatial model seems to be more accurate and scalable to model irregular network deployments. Additionally, powerful tools from stochastic geometry can be used to analytically quantify the interference and assess the performance in D2D underlaid cellular networks. 
In this D2D underlaid cellular system, we propose a decentralized opportunistic access scheme for the D2D users (transmitters) which builds on distributed SIR-based threshold scheduling and cellular exclusion regions. The main idea of our scheme is that a potential D2D link is allowed to access the cellular spectrum if the D2D transmitter is located outside the guard zones around cellular BSs (receivers) and whenever its SIR exceeds a predefined threshold.
We provide analytical expressions on the probability of successful transmission in the D2D tier and on the coverage probability in the cellular tier. Based on these analytical expressions and tight approximations, we analyze the effect of the exclusion zone radius and of the SIR threshold on network-wide key performance metrics. Furthermore, we consider the optimization problem of maximizing the area spectral efficiency of D2D communications while keeping the cellular coverage probability above a certain level.
We propose a tractable approach to solve the aforementioned optimization problem and derive in closed form the approximate optimal access probability and optimal SIR threshold. Simulation results show that our proposed opportunistic access scheme with optimized parameters provides significant performance gains in terms of D2D area spectral efficiency and cellular coverage probability as compared to state-of-the-art access control and link activation schemes.

The remainder of this paper is organized as follows. In Section~\ref{sec:sys_model}, we present the D2D underlaid cellular network model. The proposed distributed access control scheme is presented in Section~\ref{sec:scheme} and its performance is analyzed in Section~\ref{sec:perf_analysis}. The proposed scheme is optimized in Section~\ref{sec:optimization}. Simulation results are provided in Section~\ref{sec:simul} and Section~\ref{sec:conclusion} concludes our paper.

\section{System Model}
\label{sec:sys_model}
\begin{figure}
\centering
\includegraphics[scale=0.32]{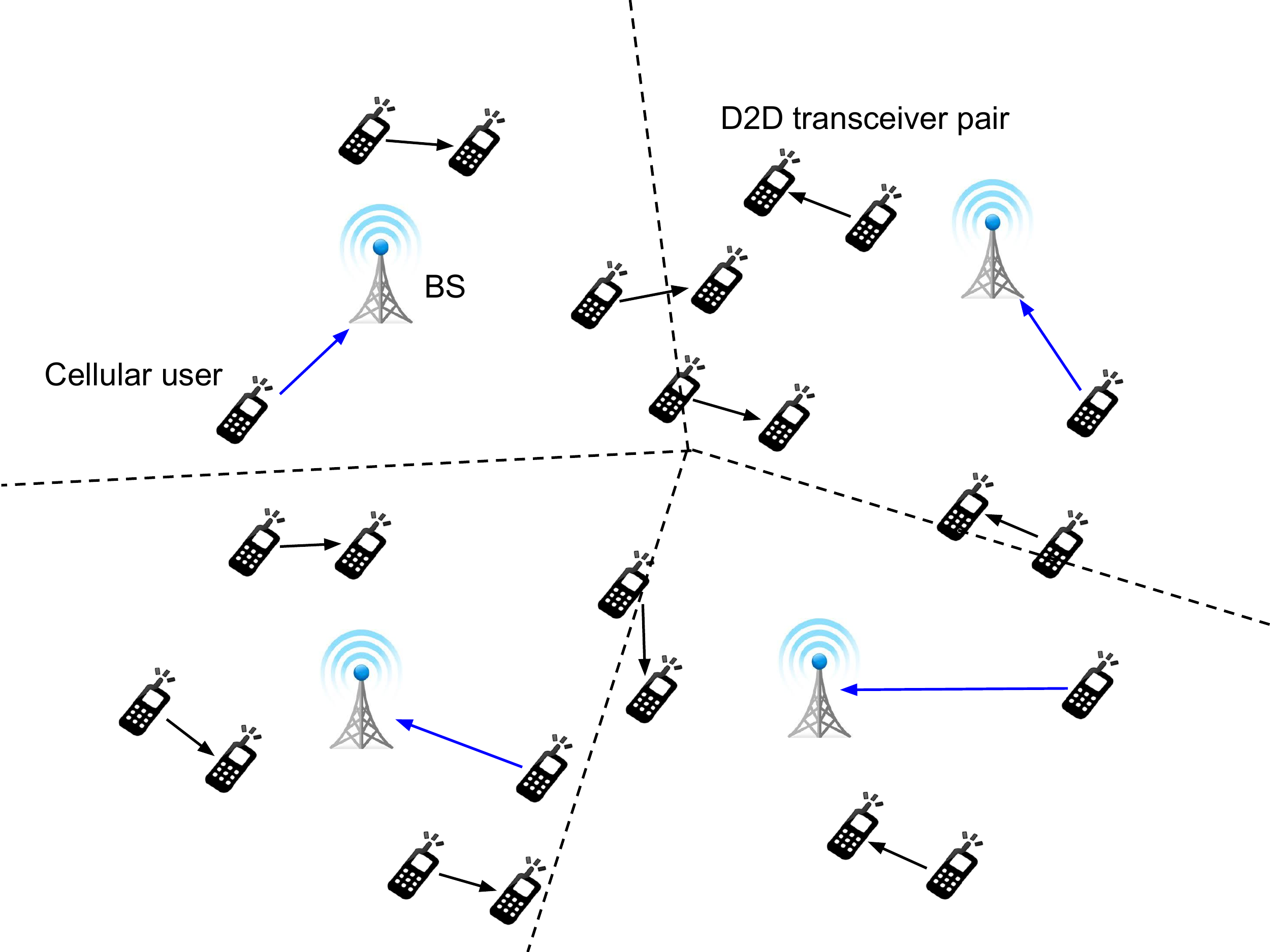}
\caption{Multi-cell D2D underlaid cellular network model. Uplink cellular users transmit to their associated BSs, while multiple D2D links access the same spectrum.}
\label{network_model}
\end{figure}

We consider a multi-cell D2D underlaid cellular network, in which D2D links share the same spectrum with cellular uplink transmissions, as shown in Fig. \ref{network_model}. The locations of cellular BSs are modeled as a homogeneous Poisson point process (PPP) $\Phi_{M}=\{ y_{i}: y_{i} \in \mathbb{R}^2 \}$ in the two-dimensional Euclidean plane $\mathbb{R}^2$ with intensity $\lambda_{M}$, where $y_i$ denotes the location of the $i$-th BS. Cellular users are placed according to some independent stationary point process and are associated to the closest base station. 
The coverage area of a BS is represented by a Poisson-Voronoi tessellation (PVT) on the plane. We assume that in each Voronoi cell there is always one active cellular user scheduled\footnote{This implies that the user density is much higher than $\lambda_M$, so that there is always at least one users to be served in the coverage region of each BS.}. Denote by $\Phi_{U}=\{ u_{i}: u_{i} \in \mathbb{R}^2 \}$ the set of active cellular uplink transmitters, since each point $u_i$ is randomly dropped in the Voronoi cell covered by $y_i$ in $\Phi_M$, the density of $\Phi_{U}$ will also be equal to $\lambda_M$.

The distribution of potential D2D transmitters follows a marked PPP $\Phi_{D}=\{ \left(x_{i}, e_{i}\right): x_{i} \in \mathbb{R}^2, e_{i} \in \{0,1\} \}$ with intensity $\lambda_D$, where $x_{i}$ denotes the location of the $i$-th D2D transmitter and $e_{i}$ denotes its transmission mode: $e_{i}=1$ means that the transmitter is active, otherwise $e_{i}=0$.  
Potential D2D receivers are distributed at random isotropic directions around their respective transmitters and at a fixed distance $d$. All base stations, uplink users and D2D nodes are equipped with a single antenna.

Due to resource sharing among the cellular uplink users and D2D pairs, the success of cellular and D2D transmissions depends on both intra-tier and cross-tier interferences. Without loss of generality, conditioning on having a D2D receiver at the origin and its associated D2D transmitter at $x_i$ with fixed distance $d$ from the receiver, its received signal-to-interference-plus-noise ratio (SINR) is given by
\begin{equation}
\text{SINR}_{i}^{\text{D}}=\frac{P_{d}|h_{i,i}|^2 d^{-\alpha}}{\!\!\sum\limits_{x_j\in \Phi_D\backslash \{x_i\}}\!\!\!\!\!\!e_{j} P_{d} |h_{j,i}|^2  d_{j,i}^{-\alpha} +\!\!\!\sum\limits_{u_k\in \Phi_U} \!\!P_{c}|h_{k,i}|^2 d_{k,i}^{-\alpha}+\sigma^{2}},
\label{sinr_d}
\end{equation}
where $P_d$ and $P_c$ denote the transmit powers of the D2D and cellular user, respectively; $h_{j,i}$ denotes the small-scale channel fading from transmitter $j$ to the $i$-th D2D receiver, with $|h_{j,i}|^2\sim \exp(1)$ (Rayleigh fading); $d_{j,i}$ denotes the distance from transmitter $j$ to $i$-th D2D receiver. We consider a distance-dependent pathloss attenuation, which follows a standard power law, i.e. $d^{-\alpha}$ where $\alpha > 2$ is the pathloss exponent; $\sigma^{2}$ denotes the background thermal noise variance. 

Similarly, for the cellular uplink communication, conditioning on having a BS at the origin and its associated cellular user at $u_i$, the received SINR at the $i$-th BS is given by
\begin{equation}
\text{SINR}_{i}^{\text{C}}=\frac{P_{c}|g_{i,i}|^2 \|u_i\|^{-\alpha}}{ \sum\limits_{x_j\in \Phi_D} \!\!\! e_{j} P_{d}|g_{j,i}|^2 l_{j,i}^{-\alpha} +\!\!\!\!\sum\limits_{u_k\in \Phi_U \backslash \{u_i\}} \!\!\!\! P_{c} |g_{k,i}|^2 l_{k,i}^{-\alpha} +\sigma^{2}},
\label{sinr_c}
\end{equation}
where $g_{j,i}$ is the channel fading from transmitter $j$ to $i$-th BS, following the same distribution as $h_{j,i}$; $l_{j,i}$ denotes the distance from transmitter $j$ to the typical BS.

In the remainder, we assume that the background noise is negligible compared to the interference, thus the SINR in \eqref{sinr_d} and \eqref{sinr_c} will be replaced by the signal-to-interference ratio (SIR) since $\sigma^{2} \to 0$. This is justified in current wireless networks, which are typically interference limited \cite{BouPanJ2009}. Background noise can be included in the subsequent analytical framework with little extra work.

\section{D2D Access Control and Link Activation Scheme}
\label{sec:scheme}
In order to alleviate the interference problem introduced by spectrum sharing in D2D underlaid cellular networks, we propose a D2D access control and link activation mechanism, which involves two main methods: (i) imposing guard zones around cellular BSs; (ii) using SIR-aware thresholding for D2D link activation; these two schemes are described below. 

\subsection{Cellular Exclusion Zones}
Any effective and reasonable design of D2D underlaid cellular networks should guarantee that devices engaging in D2D communication lie in close proximity of each other and that there is sufficient spatial separation between cellular and inband D2D transmissions. One way to achieve the latter is by creating exclusion zones around cellular users or BSs and controlling the spacing that occurs in dense network deployments. 

A first element of our proposed scheme is the use of cellular guard zones around the BSs where no D2D transmitters can lie in. By doing that, cellular uplink transmissions are protected from excessive interference due to D2D communication. This exclusion zone surrounding the macro BSs imposes that no other device is physically present. In other words, the exclusion zone creates a minimum separation among macro BSs and D2D devices.

Imposing cellular exclusion zones will create holes around the BSs. The distribution of potential D2D transmitters can then be modeled by a \textit{Poisson Hole Process} (PHP) $\Phi_{H}=\{ x_{i} \in \Phi_D:  \|x_{i}-y_{j}\|> \delta, \forall i \in \mathbb{N}_{+}, \forall j \in \mathbb{N}_{+}\}$, where $\|x_{i}-y_{j}\|$ denotes the Euclidean distance from the $i$-th D2D transmitter to the $j$-th BS, and $\delta$ denotes the exclusion zone radius. This point process model captures the spatial separation and the deactivation of D2D devices in the network in consideration. The density of $\Phi_{H}$ will then be \cite{cognitive}
\begin{equation}
	\lambda_H=\lambda_{D}\cdot\exp\left(-\lambda_M\pi \delta ^2\right).
	\label{new_density}
\end{equation}

\subsection{SIR-Aware Opportunistic Access Control}
For the potential D2D transmitters located outside the cellular exclusion zones, we propose a distributed opportunistic link scheduling protocol to determine the D2D links which are qualified to access the cellular spectrum. 

Previous work on distributed opportunistic access control has mainly focused on received signal strength (RSS) or channel-aware thresholding \cite{power_control, wcnc_opportunistic, Weber07}. Driven by the fact that local channel state information (CSI) can be obtained by sending training sequences to the receiver, the activation probability is then calculated as the probability that the RSS or SNR (channel strength) is above a certain threshold. For i.i.d. Rayleigh fading, the access probability for D2D links under threshold-based channel-aware scheduling is given by 
\begin{equation}
p_{\text{ac}}=\mathbb{P}(|h_{d}|^2 d^{-\alpha}>G_{\text{min}})=\exp(-G_{\text{min}} d^{\alpha}),
\end{equation}
where $G_{\text{min}}$ is an optimized threshold. In that case, the set of active D2D links will form a homogeneous PPP as the thresholding operation results in independent thinning of a homogeneous PPP. Existing results on the interference distribution and the outage probability in Poisson networks can be directly applied (e.g. see \cite{power_control}). The main drawback of this approach is that a potential D2D link with very strong received signal but which receives very strong interference may be activated, hence resulting in an unsuccessful transmission due to decoding errors. In other words, a D2D link with high SNR/RSS but low SINR might be activated, having marginal or even detrimental effect on the sum rate.

For that, we propose a distributed SIR-aware opportunistic access scheme that takes into account both received signal strength and interference level. A potential D2D link is allowed to access the spectrum when its SIR is above a prescribed threshold\footnote{As mentioned in the previous section, we employ SIR instead of SINR as we consider an interference-limited network.}. This scheme can use measured or estimated SIR metrics; SIR estimation can be performed prior to thresholding by allowing all D2D transmitters to transmit a test signal to their associated receivers and calculating the received SIR. Alternatively, advanced SIR estimation techniques based on sounding reference signal (SRS) can be applied. In that case, the proposed SIR-aware link activation scheme may use a two-stage protocol: in the first stage, each potential D2D link estimates its link quality (e.g. a certain function of SIR) at the beginning of each time transmission interval, and in the second stage, only the qualified D2D links, i.e. those with high estimated SIR, are active for the rest of transmission interval. Evidently, the first phase should consume a negligibly small fraction of the whole resource block. Note that the aforementioned decentralized scheduling protocol can be seen as some sort of guard zone, which extends beyond the cellular exclusion zone, where D2D devices may be present but they are deactivated by the thresholding operation. As we show below, the proposed distributed link activation scheme offers additional protection beyond that offered by the cellular exclusion zone, and - if properly optimized - it may increase the network throughput.

More formally, a potential D2D transmitter requesting access must (i) be outside the cellular guard zones and (ii) have an estimated SIR exceeding a predefined threshold. 
Let $G$ denote the SIR threshold, the transmission mode (active or not) of each potential D2D transmitter $x_i\in\Phi_D$ is
\begin{equation}
e_i=\mathds{1}\left\{\text{SIR}_{i}^{\text{D}}>G, x_i \in\Phi_H\right\}.
\end{equation}
Note that due to dependent thinning, the distribution of active D2D transmitters, denoted by $\Phi_A = \{x_{i} \in \Phi_H: \text{SIR}_{i}^{\text{D}}>G, \forall i \in \mathbb{N}_{+}\}$, is neither homogeneous PPP nor PHP.

\section{Performance Analysis}
\label{sec:perf_analysis}
The objective of this section is to investigate the effect of the design parameters of the proposed access scheme on the network performance, namely the area spectral efficiency (ASE) of the D2D network and the cellular coverage probability. In order to do so, we start from calculating the success probability of D2D links after the first step of access control, i.e. imposing cellular exclusion regions. Then, we derive the success probability of D2D links after the second step of distributed link activation with SIR-based thresholding. Finally, we study the ASE of the D2D tier and the cellular coverage probability as functions of the cellular guard zone radius and the SIR threshold, based on which we provide in Section \ref{sec:optimization} the optimal system operating parameters as a means to maximize the D2D area spectral efficiency, keeping the cellular link quality above a certain quality level. 

\subsection{Step 1: Cellular Exclusion Zones}
After the first step of the proposed scheme, the locations of potential D2D transmitters follow a PHP $\Phi_H$ with intensity given in \eqref{new_density}. The SIR distributions of the D2D and cellular links are evidently determined by the cellular exclusion zone radius. 
In this section, we analyze the success and the coverage probability in the D2D and cellular tier, respectively. 
 
\subsubsection{D2D Link Success Probability}
\label{d2d_coverage}
The D2D link success probability is defined as the probability that the SIR of a randomly chosen D2D link is higher than a prescribed SIR target $\beta$. The success probability can be seen as the fraction of D2D links having SIR higher than $\beta$ in a given realization. Building on previous analytical results for Poisson networks using stochastic geometry \cite{Baccelli, interference}, we have  
\begin{eqnarray}
p_{\text{suc}}^{\text{D}}(\beta) &=& \mathbb{E}_{0}^{!}\left[\mathbb{P}(\text{SIR}_{i}^{\text{D}}>\beta)\right] \nonumber\\
&=& \mathcal {L}_{I_{cd}} \left(\beta d^{\alpha}\frac{P_c}{P_d}\right) \mathcal{L}_{I_{dd}} \left(\beta d^{\alpha}\right), 
\label{cov_proba}
\end{eqnarray}
where $\mathbb{E}_{0}^{!}$ is the expectation with respect to the reduced Palm distribution conditioned on having the typical receiver at the origin. The terms $I_{cd}=\sum\limits_{u_k\in \Phi_U} \!\!|h_{k,i}|^2 d_{k,i}^{-\alpha}$ and $I_{dd}=\sum\limits_{x_j\in \Phi_H\backslash \{x_i\}}\!\!\!\!\!\! |h_{j,i}|^2  d_{j,i}^{-\alpha}$ denote the interference (with normalized transmit power) caused at a D2D receiver by concurrent cellular and D2D transmissions, respectively.  $\mathcal{L}_{I_{cd}}(s)=\mathbb{E}[e^{-s I_{cd} }]$ and $\mathcal{L}_{I_{dd}}(s)=\mathbb{E}[e^{-s I_{dd}}]$ are the Laplace transforms of interference $I_{cd}$ and $I_{dd}$, respectively.
Seen from the typical D2D receiver at the origin, cellular users are randomly distributed in each Voronoi cell and have the same density as the cellular BSs. Therefore, their distribution is assumed to be a homogeneous PPP, for which the Laplace transform of interference is given by \cite{Baccelli}
\begin{equation}
\mathcal {L}_{I_{cd}} \left(s\right)=\exp\left(-\frac{\pi \lambda_M s^{\frac{2}{\alpha}}}{\sinc (\frac{2}{\alpha})}\right).
\label{Lc}
\end{equation}

The exact characterization of the D2D interference and its Laplace transform $\mathcal {L}_{I_{dd}} \left(s\right)$ in a PHP is very challenging. For that, bounds and approximations have been derived in \cite{poisson_hole, cognitive}. For analytical convenience, in the following sections, we approximate $\mathcal{L}_{I_{dd}}(s)$ by its lower bound, given by\footnote{In practice, we can choose $\delta<\frac{1}{2\sqrt{\pi \lambda_M}}$ to ensure the tightness of the approximation.} 

\begin{equation}
\mathcal{L}_{I_{dd}}(s)\approx\exp \left(-\frac{\pi\lambda_D s^{\frac{2}{\alpha}}}{\sinc\left(\frac{2}{\alpha}\right)}\right),
\label{d2d_interference}
\end{equation}
which is derived using the dominant interferer approach, i.e. counting for the interferer (normally closest or strongest) whose interference contribution alone is sufficient to cause outage. In a PHP with parent process density $\lambda_D$, the distance to the nearest neighbor is very close to the one in a PPP with density $\lambda_D$.

Substituting \eqref{Lc} and \eqref{d2d_interference} into \eqref{cov_proba}, the D2D success probability becomes
\begin{equation}
p_{\text{suc}}^{\text{D}}(\beta)\approx\exp\left[-\frac{\pi d^2\beta^{\frac{2}{\alpha}}}{\sinc \left(\frac{2}{\alpha}\right)}\left(\lambda_D+ \left(\frac{P_c}{P_d}\right)^{\frac{2}{\alpha}} \lambda_M \right)\right].
\label{D2D_interference_CCDF}
\end{equation}

From the D2D success probability, we can obtain the area spectral efficiency (ASE) of the D2D network, which is the average number of successful transmissions of a certain rate that can be supported per unit area and has units of bit/s/Hz/m$^2$.

For the D2D underlaid cellular network with cellular exclusion zones, the ASE can be written as  
\begin{eqnarray}
\mathcal{T}(\beta)&=&\lambda_H \mathbb{P}(\text{SIR}_{i\in\Phi_H}^{\text{D}}>\beta)\log_{2}(1+\beta) \nonumber\\
&=&\lambda_H p_{\text{suc}}^{\text{D}}(\beta)\log_{2}(1+\beta).
\end{eqnarray}

We consider here the ASE of the D2D network as a means to quantify the benefit in terms of spatial reuse of spectrum resources with of underlaying D2D communications.
\begin{figure}
	\centering
	\includegraphics[scale=0.45]{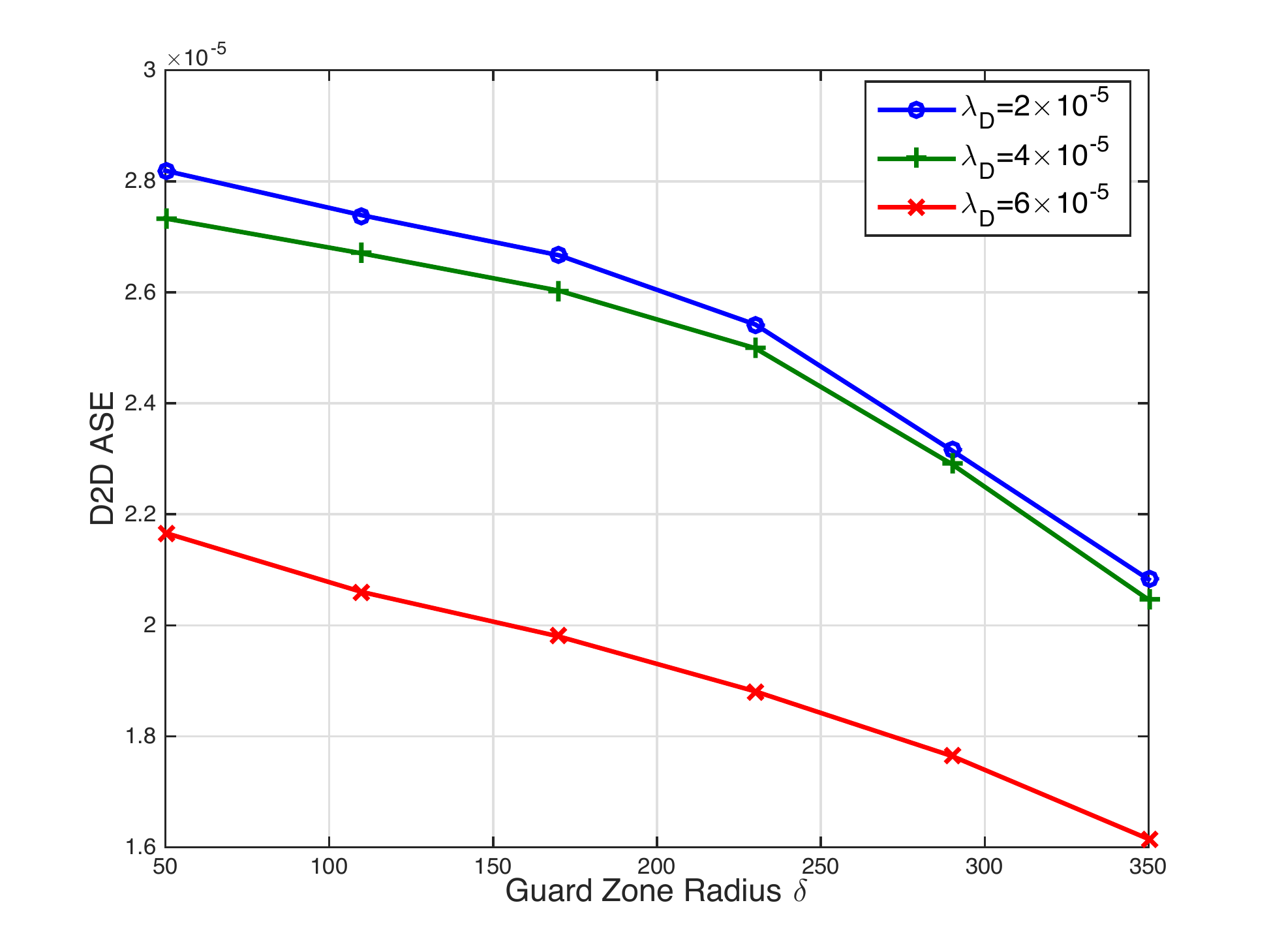}
	\caption{ASE of the D2D network vs. cellular guard zone radius $\delta$. The initial density of the potential D2D links is $\lambda_D=\{2, 4, 6\}\times10^{-5}$. }
	\label{ASE_GZ}
\end{figure}
In Fig.~\ref{ASE_GZ} we plot the ASE of the D2D network as a function of the cellular guard zone radius $\delta$. As expected, the D2D ASE is reduced when the guard zone range is increased. Interestingly though, the D2D ASE does not necessarily increase when the D2D link density augments, mainly due to the excessive interference in ultra dense D2D network deployments.

\subsubsection{Cellular Coverage Probability}
We define the cellular coverage as the probability of a random cellular link having SIR higher than a target $\gamma$, i.e. 
\begin{eqnarray}
p_{\text{cov}}^{\text{C}}(\gamma) &=& \mathbb{E}_{0}^{!}\left[\mathbb{P}(\text{SIR}_{i}^{\text{C}}>\gamma)\right] \nonumber\\
&=& \mathbb{E}_l\left[\mathcal {L}_{I_{cc}} \left(\gamma l^{\alpha}\right) \mathcal{L}_{I_{dc}} \left(\gamma l^{\alpha}\frac{P_d}{P_c}\right)\right],
\label{cov_proba_cellular}
\end{eqnarray}
where $I_{cc}=\sum\limits_{u_k\in \Phi_U \backslash \{u_i\}} \!\!\!\! |g_{k,i}|^2 l_{k,i}^{-\alpha}$ and $I_{dc}=\sum\limits_{x_j\in \Phi_H} \!\!\! |g_{j,i}|^2 l_{j,i}^{-\alpha} $ denotes the cellular interference and D2D interference to the typical cellular receiver with normalized transmit power, respectively. Assuming nearest BS association, the pdf of the cellular link distance $l$ is
\begin{equation}
f_l(x)=2\pi\lambda_M x\cdot e^{\pi\lambda_M x^2}.
\label{pdf_l}
\end{equation}

The use of cellular exclusion zones makes that the interference perceived at the typical BS does not come from the entire two-dimensional plane (whole point process of D2D transmitters). 

\begin{definition}
\label{def_1}
Consider the aggregate interference to a typical receiver at the origin $I_{\Pi}=\sum\limits_{x_{i} \in \Pi}{|h_{i}|^2 \|x_{i}\|^{-\alpha}}$, where $\Pi$ represents the spatial distribution of the interfering nodes. If $\Pi$ is generated from a homogeneous PPP with density $\lambda_{\Pi}$ and with minimum distance $r_{\textnormal{min}}$ to the typical receiver, i.e., $\|x_{i}\|\geq r_{\textnormal{min}}$, we define a modified Laplace transform of $I_{\Pi}$ as
\begin{align}
& \mathcal{L}_{I}^{1}(s, \lambda_{\Pi}, r_{\textnormal{min}}) \nonumber \\=& \mathbb{E}_{h_i, \Pi} \left[\exp \left(-s \sum_{i \in \Pi}{|h_{i}|^2 r_{i}^{-\alpha}} \right)\right]  \nonumber\\
=& \exp\left(-2\pi \lambda_{\Pi} \int_{r_{\textnormal{min}}}^{\infty}\!\!\!\left( 1-\mathbb{E}_{h_i} \left[\exp \left(-s|h_i|^2v^{-\alpha}\right)\right]\right)v \text{d} v\right) \nonumber \\
= &\exp \left(-2\pi \lambda_{\Pi} \int_{r_{\textnormal{min}}}^{\infty} \frac{s v^{-\alpha}}{1+s v^{-\alpha}} v \text{d} v\right) .
\label{laplace_tranform}
\end{align}
\end{definition} 

From \cite{cellular_uplink} we have that the interfering uplink users can be modeled by a softcore process due to the pairwise correlation among active cellular users in a given time slot. Due to the intractability of the aforementioned process, the received interference can be approximated as coming from PPP-distributed interfering nodes outside the circle centered at the typical BS with the same area as its Voronoi cell. 

\begin{propi}
	\label{propo_laplace}
	The Laplace transform of interference $I_{cc}$ can be approximated by 
	\begin{equation}
	\label{uplink_Icc}
	\mathcal {L}_{I_{cc}} (s)\approx \mathcal{L}_{I}^{1}(s, \lambda_M, d_{\min})
	\end{equation}
	 where the pdf of $d_{\min}$ is given by
	\begin{equation}
	f_{d_{\min}}(r)= 2\frac{(3.5\pi \lambda_M)^{3.5}}{\Gamma(3.5)} r^{6} \exp(-3.5\pi \lambda_M r^2).
	\label{d_min approxi}
	\end{equation}
\end{propi}
\begin{IEEEproof}
	\textnormal{See Appendix \ref{appen_laplace}.}
\end{IEEEproof}

\begin{figure}
	\centering
	\includegraphics[scale=0.45]{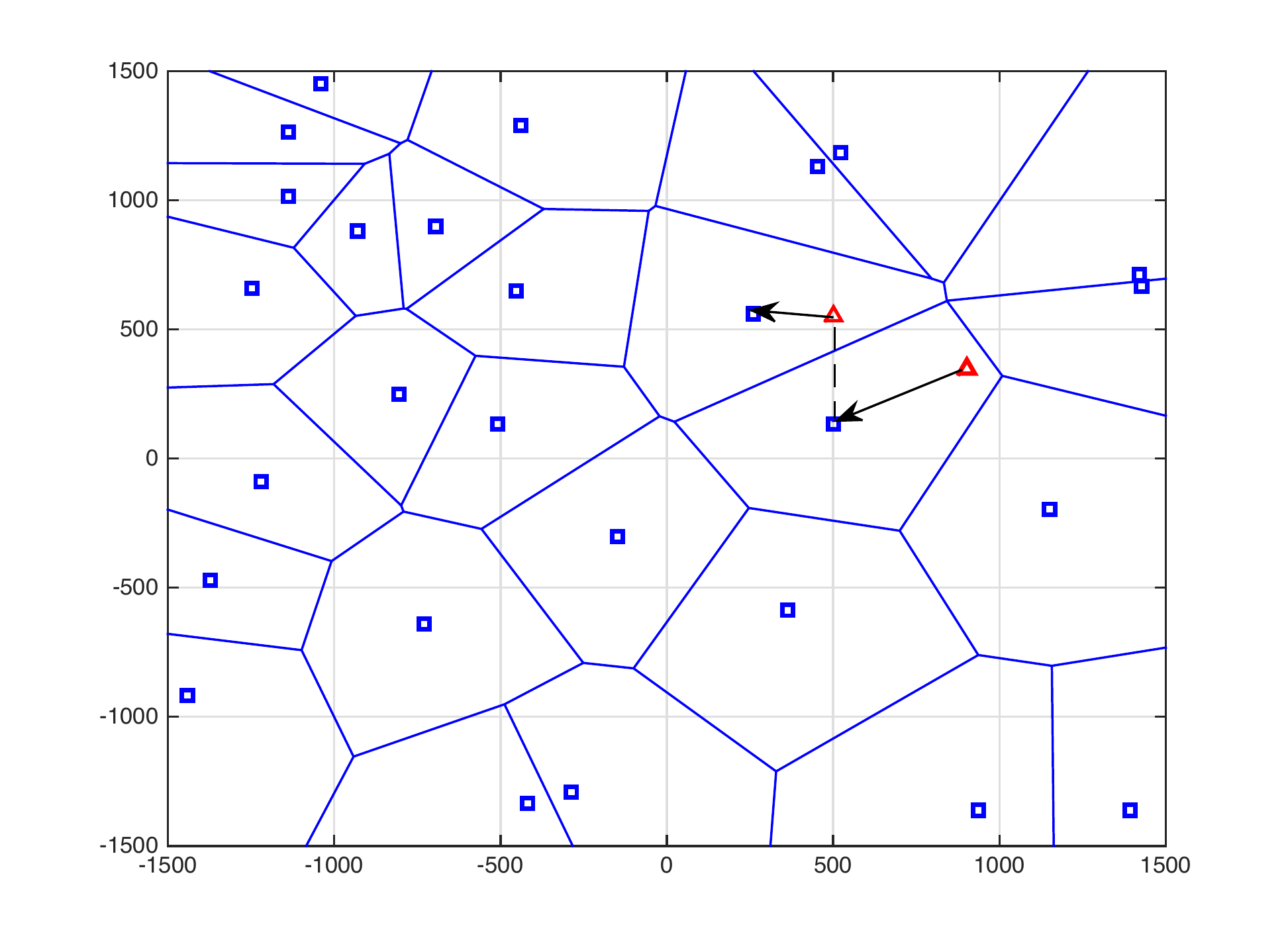}
	\caption{Voronoi tessellation of a cellular network with BSs distributed according to a homogeneous PPP. BSs are represented by blue squares. Red triangles represent two random users served by two BSs nearby. }
	\label{voronoi}
\end{figure}

\begin{figure}
	\centering
	\includegraphics[scale=0.45]{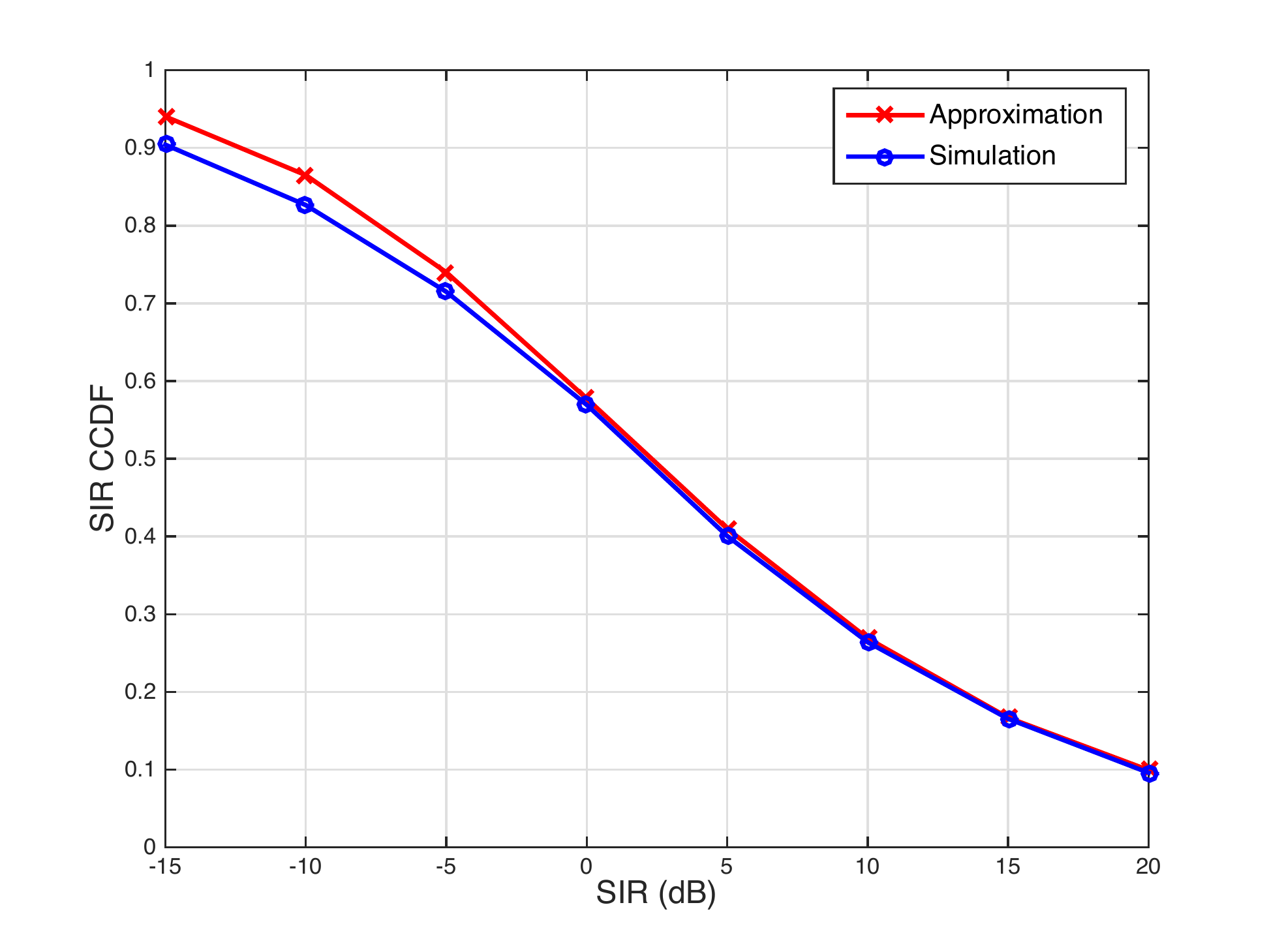}
	\caption{SIR CCDF at cellular receivers (BSs) when only considering cellular interference $I_{cc}$. BS density $\lambda_M=10^{-6}$. Each BS serves one cellular user in its Voronoi cell at a given time.}
	\label{cellular_ccdf}
\end{figure}

Fig. \ref{cellular_ccdf} compares the simulated and theoretical SIR CCDF of cellular uplink users while considering only cellular interference $I_{cc}$. We can see that our approximation in Proposition~\ref{propo_laplace} gives relatively accurate result in terms of the SIR distribution in uplink cellular networks .

\begin{remark}
	According to the nearest BS association, each uplink user is uniformly distributed in the Voronoi cell of its connected BS. It is worth noticing that the user being connected to the nearest BS is not equivalent to that the BS is associated to the nearest user, as assumed in \cite{uplinkmodeling}. The nearest interfering uplink user might be closer to the typical BS than its own tagged user (see Fig. \ref{voronoi}). The above proposition captures the discrepancy between these two association conditions and provides a tight approximation for the uplink cellular coverage probability. 
\end{remark}

As for the Laplace transform of D2D interference $\mathcal {L}_{I_d}$, there also exists a minimum distance $\delta$ between the nearest D2D transmitter and the typical BS. Since $\Phi_H$ is generated from a PPP with intensity $\lambda_D$ and with guard zone size $\delta$, we have 
\begin{equation}
\mathcal {L}_{I_{dc}} (s) \approx\mathcal{L}_{I}^{1}(s, \lambda_D, \delta).
\label{L_Idc}
\end{equation}
Note that this approximation is actually a lower bound on the actual Laplace transform of $I_{dc}$ since the density of D2D users in $\Phi_H$ has lower density than $\lambda_D$. Tighter approximation on the SIR distribution in a PHP can be found in \cite{poisson_hole}, however, its involved expression would only make the related analysis more cumbersome. For that, we adhere to \eqref{L_Idc} as a baseline approximation, which will be shown to have tolerable error gap in the optimization problem we study in Section~\ref{sec:optimization}. 

Substituting \eqref{L_Idc} and the approximation of $\mathcal {L}_{I_{cc}} (s)$ proposed in Proposition \ref{propo_laplace} into \eqref{cov_proba_cellular}, we obtain the cellular coverage probability as follows.

\begin{propi}
\label{theorem_cellular}
The cellular link coverage probability in a D2D underlaid cellular network with guard zones of radius $\delta$ around the BSs is given by
\begin{equation}
\label{cov_cellular}
\begin{split}
 p^{C}_{\textnormal{cov}}(\gamma)\approx\int_{0}^{\infty}  f_l(x) \int_{0}^{\infty} f_{d_{\min}} (r) \mathcal {L}_{I}^{1} \left(\gamma x^{\alpha}, \lambda_M, r\right)  \\ \mathcal{L}_{I}^{1}(\gamma x^{\alpha}\frac{P_d}{P_c}, \lambda_{\text{D}}, \delta) \textnormal{d}r \textnormal{d}x,
 \end{split}
\end{equation}
\textnormal{where $
	f_{d_{\min}}(r)=2\frac{(3.5\pi \lambda_M)^{3.5}}{\Gamma(3.5)} r^{6} \exp(-3.5\pi \lambda_M r^2)$, $f_l(x)=2\pi\lambda_M x\cdot e^{\pi\lambda_M x^2}$}. 
\end{propi}

\begin{figure}
	\centering
	\includegraphics[scale=0.45]{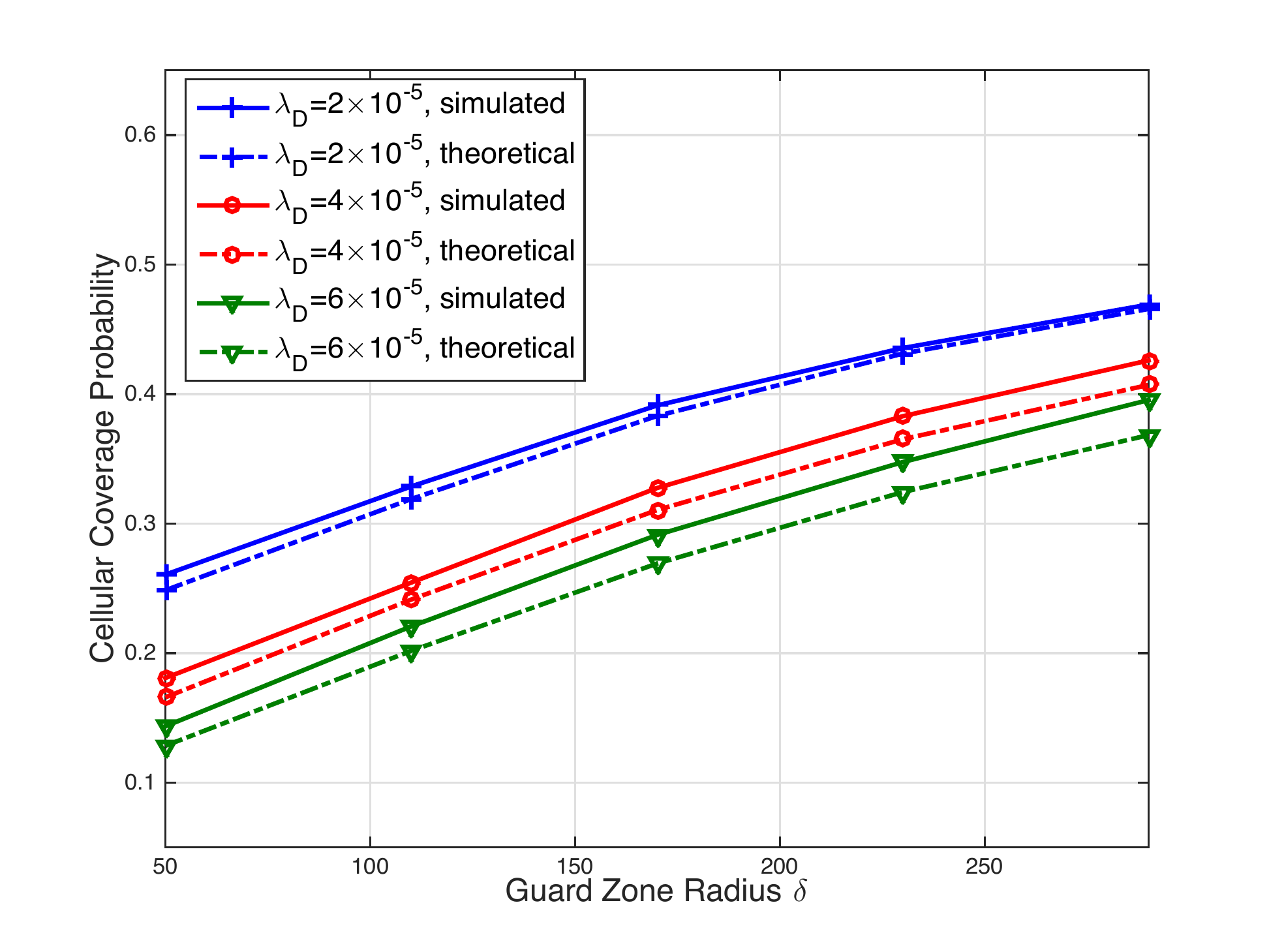}
	\caption{Cellular coverage probability vs. guard zone radius $\delta$. The initial density of the potential D2D links is $\lambda_D=\{2, 4, 6\}\times10^{-5}$ /m$^2$. Other system parameters are as in Table~\ref{system_params}.}
	\label{cellular_cov}
\end{figure}

In Fig.~\ref{cellular_cov} we compare the simulated cellular coverage probability with the ones obtained from Proposition~\ref{theorem_cellular}. We see that the theoretical result with $\lambda_D=2\times10^{-5}$ is quite close to the simulated one, while the approximation error becomes larger when $\lambda_D$ increases, which is mainly due to the approximation in \eqref{L_Idc}. Combined with Fig.~\ref{ASE_GZ}, we conclude that increasing the guard zone radius $\delta$ eliminates the potential improvement in terms of D2D area spectral efficiency, but offers better protection to the cellular users. Moreover, we see that setting guard zones alone is not efficient in D2D underlaid cellular networks. This motivates us to introducing a second step in our proposed access control scheme, that of using SIR-aware D2D link activation in order to achieve the highest possible D2D ASE for any value of D2D link density.

\subsection{Step 2: SIR-aware Opportunistic Access}
Denoting the set of active D2D transmitters by $\Phi_A=\{x_{i} \in \Phi_H: \text{SIR}_{i}^{\text{D}}>G, \forall i \in \mathbb{N}_{+}\}$ with average density $\lambda_A$, the success probability of a typical active D2D link is a conditional probability given that the $i$-th D2D pair could be active, i.e. the transmitter does not fall within the cellular exclusion zones and its estimated $\text{SIR}_{i}$ exceeds the threshold $G$. Given the SIR threshold $\beta$ for successful D2D transmission, the (conditional) success probability is $\mathbb{P}(\text{SIR}_{i\in\Phi_A}>\beta|\text{SIR}_{i\in \Phi_H}>G)$.
The ASE of the D2D network can be expressed as  
\begin{equation}
\mathcal{T}_{D}(\beta)=\lambda_A \mathbb{P}(\text{SIR}_{i\in\Phi_A}>\beta|\text{SIR}_{i\in \Phi_H}>G)\log_{2}(1+\beta).
\end{equation}

From our analysis in Section~\ref{d2d_coverage}, for the potential D2D transmitters in $\Phi_H$, the D2D access (activation) probability $p_{s}$ is the same as the D2D link success probability with $G$ as the SIR target. From \eqref{D2D_interference_CCDF}, we have
\begin{eqnarray}
p_{s}&=&\mathbb{P}[\text{SIR}_{i\in \Phi_H}^{\text{D}}>G] \nonumber \\
&\approx&\exp\left[-\frac{\pi d^2 G^{\frac{2}{\alpha}}}{\sinc \left(\frac{2}{\alpha}\right)}\left(\lambda_D+ \left(\frac{P_c}{P_d}\right)^{\frac{2}{\alpha}}\!\!\! \lambda_M \right)\right].
\label{Ps}
\end{eqnarray}

Note that $p_{s}$ is a mean value by averaging over the fading statistics and all realizations of PHP $\Phi_H$. For a specific PHP realization or conditioned on $\Phi_H$, each D2D link experiences different SIR and thus should in principle be configured with different access probability depending on its location and surroundings, i.e. the locations of nearby D2D transmitters in this realization. In other words, when there are many interfering nodes in the vicinity of this D2D link, this link has lower access probability than a link in an area isolated from nearby interferers due to the fact that it has potentially lower SIR. So for each realization of $\Phi_H$, $p_{s}$ actually represents the proportion of D2D links that are allowed to access the spectrum. 

Applying the proposed SIR-aware opportunistic access control results in dependent thinning of the PHP $\Phi_H$, thus the set of active D2D transmitters $\Phi_{A}$ is hard or impossible to define (it is neither PPP nor PHP). For that, we resort to the approximation that $\Phi_{A}$ is a PHP with intensity given by
\begin{equation}
\lambda_A\simeq p_s\lambda_H= p_s\lambda_D \cdot \exp(-\lambda_M\pi\delta^2).
\label{lambda_A}
\end{equation}
Rewriting the ASE of the D2D network as a function of the guard zone radius $\delta$ and D2D average access probability $p_s$, we obtain
\begin{equation}
\begin{split}
\mathcal{T}_{D}(\delta, p_s)\simeq p_s\lambda_D \cdot \exp(-\lambda_M\pi\delta^2)\log_{2}(1+\beta)  \\ \cdot\mathbb{P}(\text{SIR}_{i\in\Phi_A}>\beta|\text{SIR}_{i\in \Phi_H}>G).
\end{split}
\label{throughput_definition}
\end{equation}

From Proposition~\ref{theorem_cellular}, the cellular coverage probability when the locations of active D2D links follow $\Phi_A$ with intensity $\lambda_A$ is given by
\begin{equation}
\begin{split}
p^{C}_{\textnormal{cov}}(\gamma)\approx\int_{0}^{\infty}  f_l(x) \int_{0}^{\infty} f_{d_{\min}} (r) \mathcal {L}_{I}^{1} \left(\gamma x^{\alpha}, \lambda_M, r\right)  \\ \mathcal{L}_{I}^{1}(\gamma x^{\alpha}\frac{P_d}{P_c}, \lambda_{D}p_s, \delta) \textnormal{d}r \textnormal{d}x.
\end{split}
\end{equation}

\begin{figure}[!ht]
	\centering
	\subfigure[ASE of D2D network vs. $(\delta, p_s)$ ]{
		\includegraphics[scale=0.45]{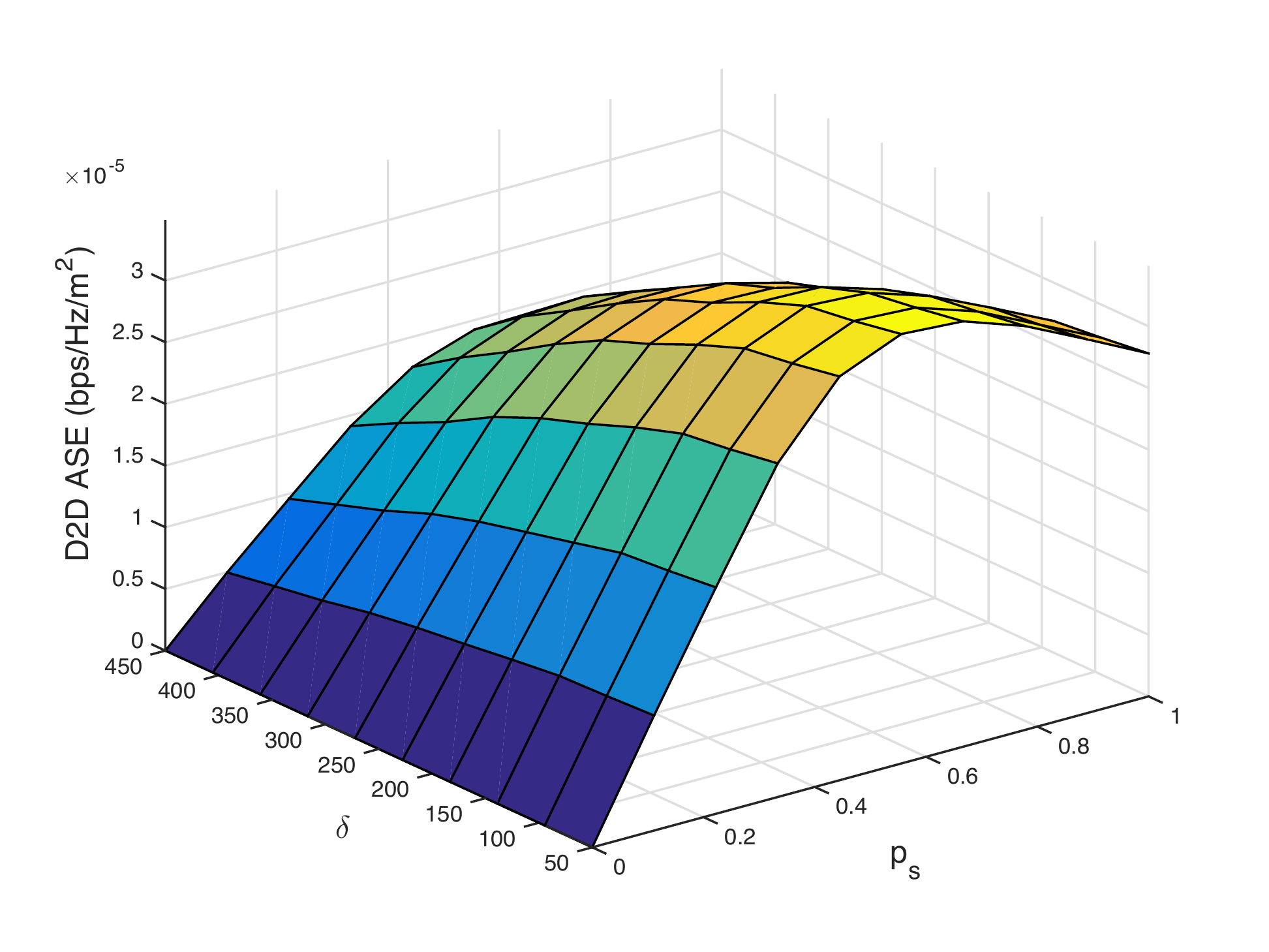}
		\label{ase_3d}
	}
	\subfigure[Cellular coverage probability vs. $(\delta, p_s)$]{
		\includegraphics[scale=0.45]{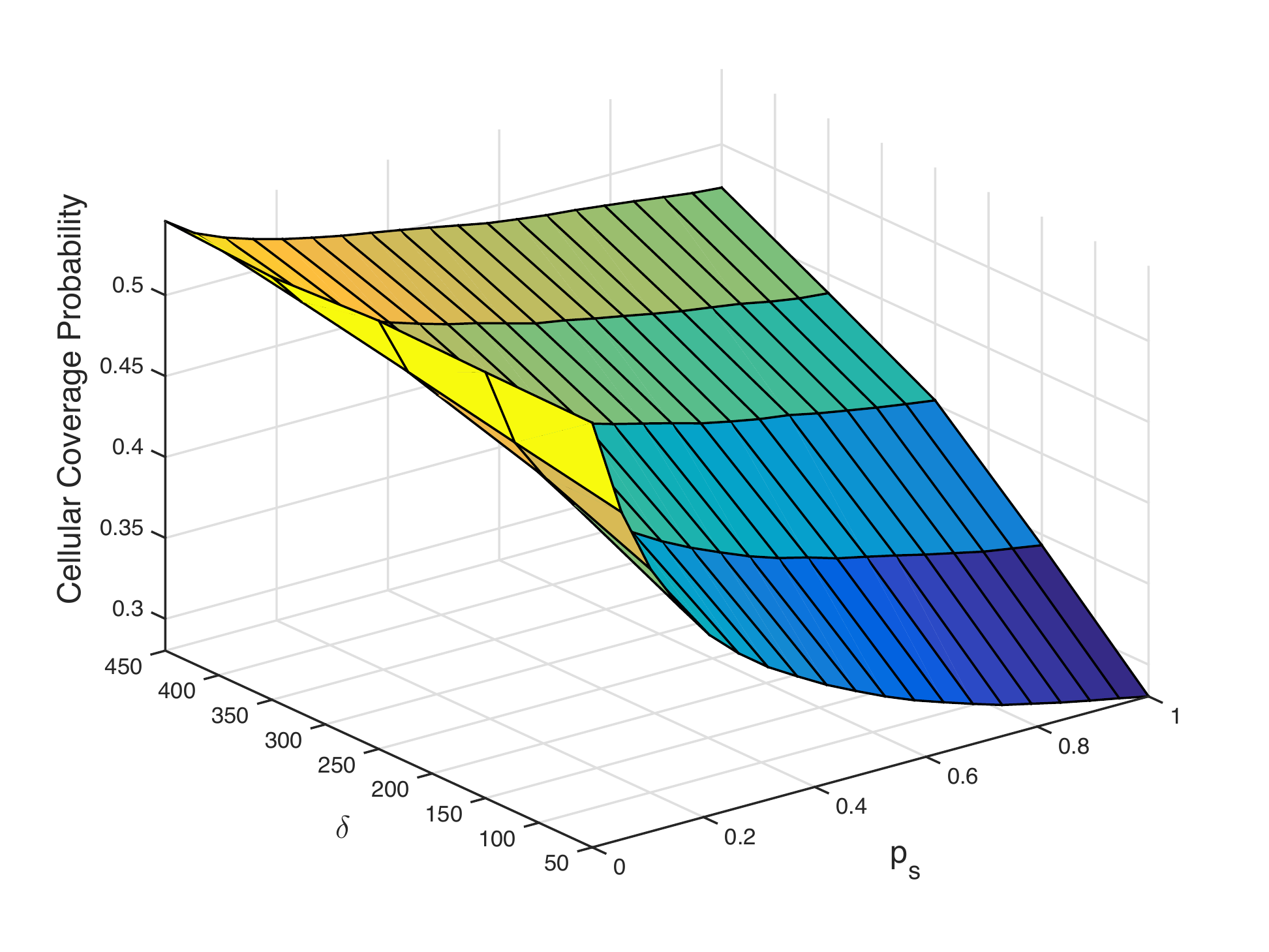}
		\label{coverage_3d}
	}
\end{figure}

In order to understand how $\delta$ and $p_s$ affect the network performance, we plot the ASE of the D2D tier and the cellular coverage probability in Fig. \ref{ase_3d} and Fig. \ref{coverage_3d}, respectively. The density of the parent process (initial D2D density) is $\lambda_D=4\times10^{-5}$ /m$^2$ and all other parameters are set as in Table~\ref{system_params}. The active D2D transmitters are selected by sorting the estimated SIR value of all potential D2D links in $\Phi_H$ and choosing the $p_s$ percentage of D2D links with highest estimated SIR. From Fig. \ref{ase_3d} we observe that larger $\delta$ leads to lower D2D ASE. For a given value of $\delta$, there always exists an optimal $p_s$ for which the D2D underlaid network maximizes its ASE. 
As for the cellular coverage, from Fig. \ref{coverage_3d}, expectedly, $p_{\text{cov}}^{\text{C}}$ increases with $\delta$ and decreases with $p_s$ due to interference reduction.
Combining together these two figures, we can easily understand that there exists an optimal point for which the D2D ASE is maximized while guaranteeing that the cellular coverage probability is above a certain threshold, if $p_s$ and $\delta$ are properly tuned.

\section{Distributed Opportunistic Access Optimization}
\label{sec:optimization}
In this section, we aim at optimizing the two key operating parameters of the proposed opportunistic access scheme in order to maximize the D2D area spectral efficiency while keeping the cellular link quality above a certain level. The optimization problem is cast as follows
\begin{equation}
\centering
(\delta^*, p_s^*)=\argmax_{(\delta, p_s) } ~ \mathcal{T}_D,
\end{equation}
subject to
\begin{eqnarray}
\label{condition 1} \delta& \in& [0,\infty],  \nonumber\\
\label{condition 2}  p_s &\in& [0,1], \nonumber\\
\label{condition 3}  p_{\text{cov}}^{\text{C}} &\geq& (1-\mu) p_{\max}^{\text{C}} ,
\end{eqnarray} 
where $\mu\in[0,1]$ is the maximum coverage degradation coefficient, and $p_{\max}^{\text{C}}$ is the cellular coverage probability without D2D interference (single-tier network).
From Proposition \ref{theorem_cellular}, when $\lambda_{H}=0$, we have
\begin{equation}
p_{\max}^{\text{C}}\approx\int_{0}^{\infty}  f_l(x) \int_{0}^{\infty} f_{d_{\min}} (r) \mathcal {L}_{I}^{1} \left(\gamma x^{\alpha}, \lambda_M, r\right) \textnormal{d}r \textnormal{d}x.
\end{equation}
Then the condition in \eqref{condition 3} can be rewritten as 
\begin{equation}
\begin{split}
\int_{0}^{\infty} \!\!\! \!\!f_l(x) \int_{0}^{\infty} \!\!\!\!\!f_{d_{\min}} (r) \mathcal {L}_{I}^{1} \left(\gamma x^{\alpha}, \lambda_M, r\right) \mathcal{L}_{I}^{1}(\gamma x^{\alpha}\frac{P_d}{P_c}, p_s \lambda_{D}, \delta) \textnormal{d}r \textnormal{d}x \\\geq (1-\mu )\int_{0}^{\infty} \!\!\!\!f_l(x) \int_{0}^{\infty}\!\!\!\! f_{d_{\min}} (r) \mathcal {L}_{I}^{1} \left(\gamma x^{\alpha}, \lambda_M, r\right) \textnormal{d}r \textnormal{d}x.
\end{split}
\label{inequality_cellular}
\end{equation}

\subsection{Decoupled Optimization}
A joint design of $\delta$ and $p_s$ seems cumbersome to be obtained, mainly due to the involved expressions for the coverage probability and the area spectral efficiency.  
In order to solve the above optimization problem, we take on a decoupled approach and proceed with the following procedure:
\begin{enumerate}
	\item For a random value of $\delta$, search for 
\begin{equation}
\centering
p_s^* (\delta)=\argmax_{p_s \in[0,1]} ~ \mathcal{T}_D (p_s, \delta),
\label{new_objective}
\end{equation}
where $\mathcal{T}_D (p_s, \delta)=p_s\lambda_{H}\log_{2}(1+\beta) \mathbb{P}(\text{SIR}_{i\in\Phi_A}\!\!>\!\beta | \text{SIR}_{i\in \Phi_H}\!\!>\!G)$ with $\lambda_{H}=\lambda_D \cdot \exp(-\lambda_M\pi\delta^2)$.
\item Replace $p_s$ in \eqref{inequality_cellular} by the $p_s^{\star}(\delta)$ obtained in the first step, calculate numerically the minimum guard zone radius $\delta^{\star}$ by solving the following equation
\begin{equation}
p_{\text{cov}}^{\text{C}} (\delta^{\star}, p_s^{\star}) = (1-\mu) p_{\max}^{\text{C}}.
\label{optimal_delta_condition}
\end{equation} 
\item Substitute the value of $\delta^{\star}$ in \eqref{new_objective} and obtain the optimized access probability $p_s^* (\delta^{\star})$. 
\end{enumerate}

The values of $(\delta^{\star}, p_s^{\star})$ solving the decoupled optimization problem are clearly not optimal; however, our simulation results provided in the following section show that the solutions of the decoupled approach are very close to the optimal solution of the joint optimization. In the remainder of this section, we focus on deriving the optimal access probability $p_s^{\star}$ as the solution to \eqref{new_objective}, as well as the optimal SIR threshold $G^{\star}$ according to the relation between $G^{\star}$ and  $p_s^{\star}$ given in \eqref{Ps}. 

\subsection{SIR Threshold Optimization for Given $\delta$}
\label{section_conditional}
From the definition of the D2D ASE given in \eqref{throughput_definition}, we see that the conditional D2D success probability $\mathbb{P}(\text{SIR}_{i\in\Phi_A}>\beta, \text{SIR}_{i\in \Phi_H}>G)$ concerns two dependent events. A potential D2D link with high SIR during the first stage of our SIR-aware protocol is very likely to have high SIR once allowed to be active. Although it seems hard or impossible to obtain a neat expression for the conditional probability, we approximate the optimal access probability $p_{s}$ as the crossing point between the following two regimes: 
\begin{itemize}
\item if $G\gg\beta$, which implies $p_{s}\rightarrow 0$, the set of nodes $\mathcal{A}=\left\{i\in\Phi_H: \text{SIR}_{i}>G \right\}$ can be approximately seen as a subset of $\mathcal{B}=\left\{i\in\Phi_A : \text{SIR}_{i}>\beta \right\}$, thus\\
\begin{equation}
\mathbb{P}(\text{SIR}_{i\in\Phi_A}>\beta| \text{SIR}_{i\in \Phi_H}>G ) 
\simeq 1
\label{extreme1}
\end{equation}

\item if $G\ll\beta$, which implies $p_{s}\rightarrow 1$, the set of nodes $\mathcal{B}=\left\{i\in\Phi_A : \text{SIR}_{i}>\beta \right\}$ can be approximately seen as a subset of $\mathcal{A}=\left\{i\in\Phi_H : \text{SIR}_{i}>G \right\}$, thus
\begin{align}
&\mathbb{P}(\text{SIR}_{i\in\Phi_A}>\beta| \text{SIR}_{i\in \Phi_H}>G) 
\simeq \frac{\mathbb{P}(\text{SIR}_{i\in\Phi_A}>\beta) }{\mathbb{P}(\text{SIR}_{i\in \Phi_H}>G)} \nonumber\\ 
=& \frac{1}{p_s} \exp\left[-\frac{\pi d^2\beta^{\frac{2}{\alpha}}}{\sinc \left(\frac{2}{\alpha}\right)}\left(p_s\lambda_D+ \left(P_c/P_d\right)^{\frac{2}{\alpha}}\lambda_M \right)\right].
\label{extreme2}
\end{align}

\end{itemize}
Therefore, the ASE of the D2D tier is written as a function of $p_{s}$ as
\begin{equation}
\mathcal{T}_D(p_s) = 
\left\lbrace 
\begin{array}{ccc}
   \lambda_H p_{s}\log_{2}(1+\beta)
    &  p_{s}\rightarrow 0\\
    && \\
   \lambda_H e^{-\xi \beta^{\frac{2}{\alpha}}\left(p_s\lambda_D+ \kappa \lambda_M \right)}\log_{2}(1+\beta) 
    &  p_{s}\rightarrow 1,\\
\end{array} \right.
\end{equation} 
where $\xi=\frac{\pi d^2}{\sinc \left(\frac{2}{\alpha}\right)}$ and $\kappa=\left(\frac{P_c}{P_d}\right)^{\frac{2}{\alpha}}$.

The approximately optimal access probability $p_s^{\star}$ and the approximately optimal SIR threshold $G^{\star}$ are given in the following proposition: 
\begin{propi}
\label{prop2}
The approximately optimal access probability for the proposed SIR-aware opportunistic access scheme (based on the conditional D2D success probability) is given by
\begin{equation}
p_{s}^{\star} \simeq \min\left\{\frac{\mathcal{W}\left(\lambda_D \xi \beta^{\frac{2}{\alpha}}  e^{-\kappa \lambda_M \xi \beta^{\frac{2}{\alpha}}}\right)}{\lambda_D \xi \beta^{\frac{2}{\alpha}}},1\right\},
\end{equation}
where $\mathcal{W}$ denotes Lambert W function.
The optimal SIR threshold in this case is approximately given as
\begin{equation}
G^{\star}\simeq\left[\frac{-\ln p_{s}^{\star} }{\xi(\lambda_D+\kappa \lambda_M)}\right]^{\frac{\alpha}{2}},
\label{optimal_SIR}
\end{equation}
where $\xi=\frac{\pi d^2}{\sinc \left(\frac{2}{\alpha}\right)}$ and $\kappa=\left(\frac{P_c}{P_d}\right)^{\frac{2}{\alpha}}$.
\end{propi}
\begin{IEEEproof}
	\textnormal{See Appendix \ref{appen3}.}
\end{IEEEproof}

\begin{figure}
	\centering
	\includegraphics[scale=0.45]{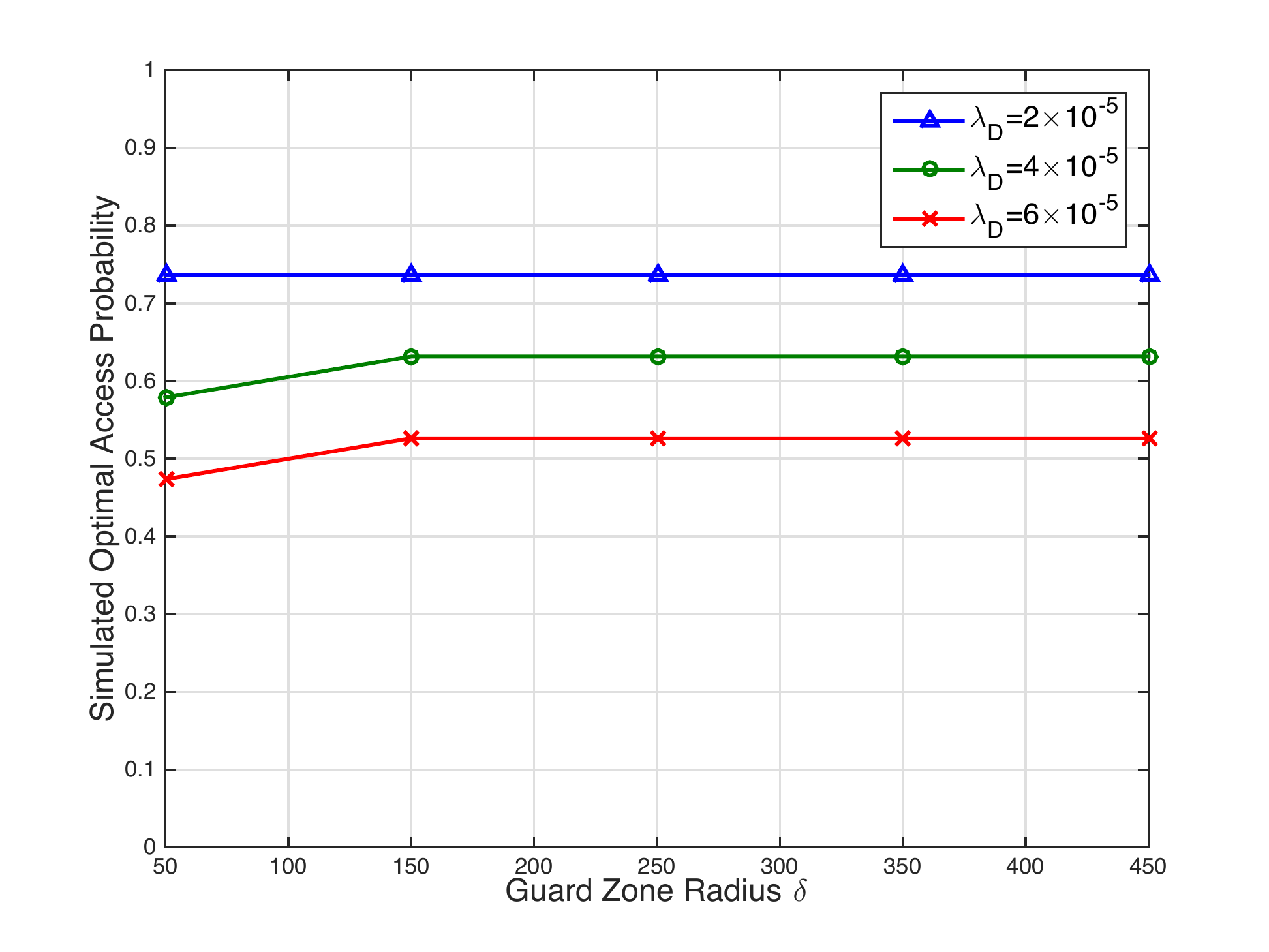}
	\caption{Simulated optimal access probability that gives the highest D2D ASE for different cellular guard zone radius $\delta$. In the simulations, only the $p_s$ percentage of D2D links with the highest estimated SIR are allowed to be active. The optimal value of $p_s$ are obtained through exhaustive search.}
	\label{optimal_ps_vs_delta}
\end{figure}

\begin{remark}
	The derived $p_s^{\star}$ is independent of $\delta$ because of the approximation used in \eqref{D2D_interference_CCDF} when the exclusion zones do not overlap, i.e. $\delta<\frac{1}{2\sqrt{\pi \lambda_M}}$. The cellular guard zone range affects only the density of active D2D links and has little impact on the optimal D2D success probability. In order to validate this assumption, we plot in Fig. \ref{optimal_ps_vs_delta} the simulated optimal $p_s$ obtained by exhaustive search that satisfies \eqref{new_objective} for different values of $\delta$. It evinces that the optimal access probability in terms of D2D ASE maximization is not very sensitive to cellular guard zone radius $\delta$. Hence, the decoupled optimization may give approximately optimal values of $p_s$ and $\delta$ if properly performed.
\end{remark}

\section{Simulation Results}
\label{sec:simul}
\begin{figure}
	\centering
	\includegraphics[scale=0.6]{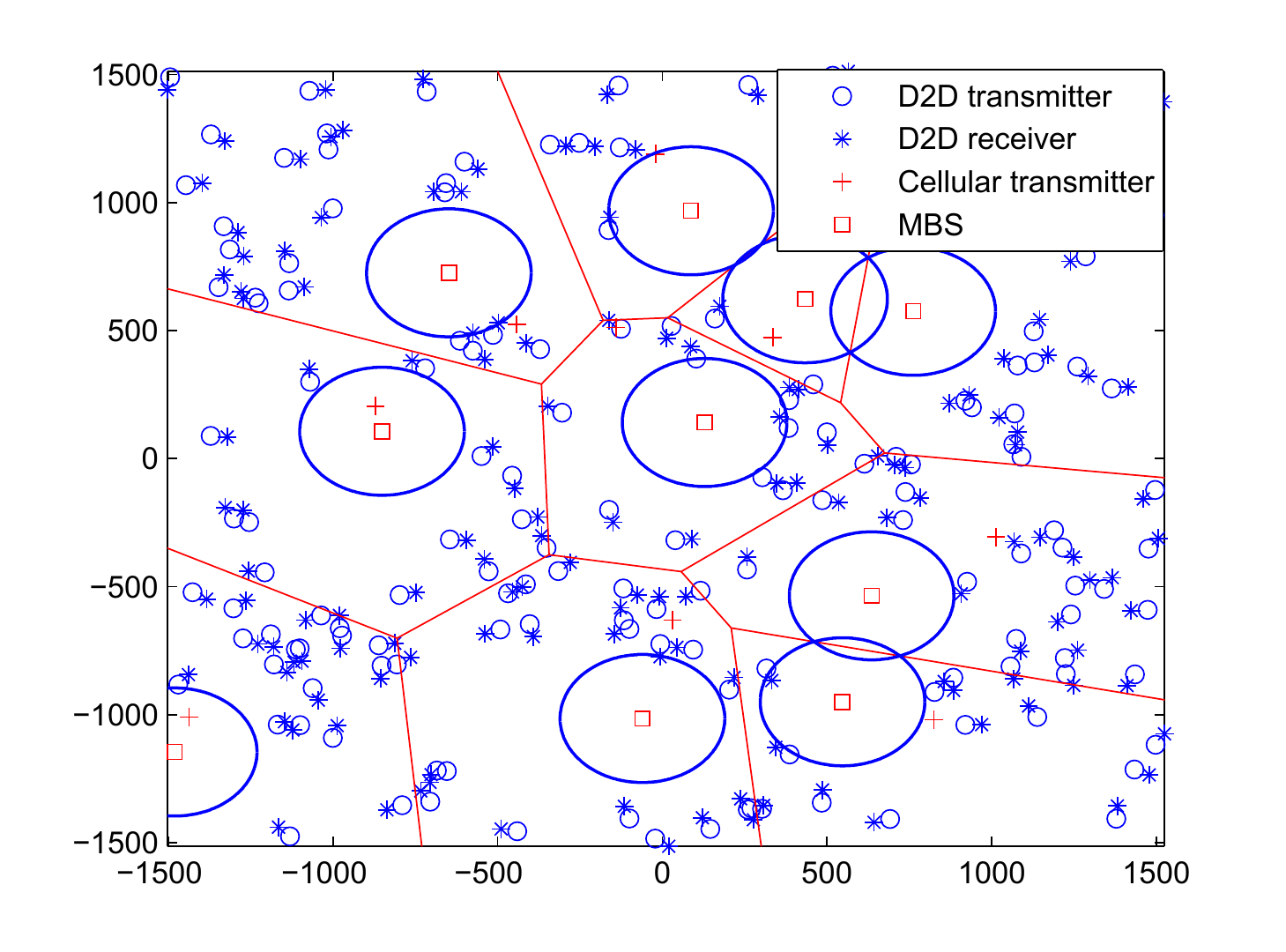}
	\caption{A snapshot of a multi-cell D2D underlaid cellular network with cellular guard zones around macrocell BSs. Potential D2D link density $\lambda_D=2\times 10^{-5}$. Only one cellular user in each Voronoi cell is active at a time. }
	\label{link_geometry}
\end{figure}

In this section, we assess the performance of the proposed access control scheme for D2D underlaid cellular networks. 
Simulations are performed on a square region of surface $3000\times3000~\text{m}^2$. Both cellular BSs and potential D2D transmitters are distributed according to  a homogeneous PPP with intensity $\lambda_M$ and $\lambda_D$, respectively. The uplink users are uniformly distributed in each Voronoi cell covered by the nearest BS. Each D2D receiver is placed at a random direction around its transmitter with a fixed distance $d$. Fig. \ref{link_geometry} shows a snapshot of the network layout with $\lambda_D=2\times 10^{-5}$ /m$^2$ and with cellular guard zone radius $\delta=250$ m. Rayleigh fading is considered for both cellular and D2D links with $\mathbb{E}[|h|^2]=1$. All other parameters are set according to Table~\ref{system_params}.

All results are obtained by averaging over $4000$ realizations. 
The following access strategies are also simulated for comparison and for evincing the performance gains of the proposed scheme:
\begin{itemize}
	\item Only guard zone (GZ) scheme: all potential D2D links outside the cellular guard zones in $\Phi_H$ are active. The guard zone radius $\delta$ is chosen to satisfy the cellular coverage constraints. This basically corresponds to the first step of our proposed access scheme.
	\item Channel-aware access control (AC) with cellular guard zones: the link activation scheme in \cite{power_control} is applied together with cellular guard zones that satisfy the cellular coverage constraints.
\end{itemize}

\begin{table}[t]
  \centering
  \caption{Simulation Setup}
  \renewcommand{\arraystretch}{1.2}
  \begin{tabular}{c|c}
  \firsthline
  \textbf{Parameters}              & \textbf{Values}  \\
  \hline   
 \small {Macrocell BS density ($\lambda_M$) }& \small {$10^{-6} $  }    \\
  \small {D2D link density ($\lambda_D$)   }  & \small {$[2, 10]\times10^{-5}$ /m$^2$}\\
  \small {D2D link length ($d$)       } & \small {50 (m)  }    \\
  \small {Pathloss exponent ($\alpha$) }    & \small {4  }   \\  
  \small {D2D SIR threshold ($\beta$)} & \small {$5$ dB  }   \\  
  \small {Cellular SIR threshold ($\gamma$) }& \small {$0$ dB  }   \\  
  \small {Cellular user transmit power ($P_{c}$)} & \small {10 (mW)  }    \\
  \small {D2D user transmit power ($P_{d}$)   }   & \small {0.1 (mW)  }    \\
 \small { Cellular degradation coefficient ($\mu$)}  & \small {$30\%$ }    \\
  \lasthline
  \end{tabular}
  \label{system_params}
\end{table}

\subsection{Proposed Access Control with Optimized $(p_s, \delta)$}
\begin{figure}
	\centering
	\includegraphics[scale=0.45]{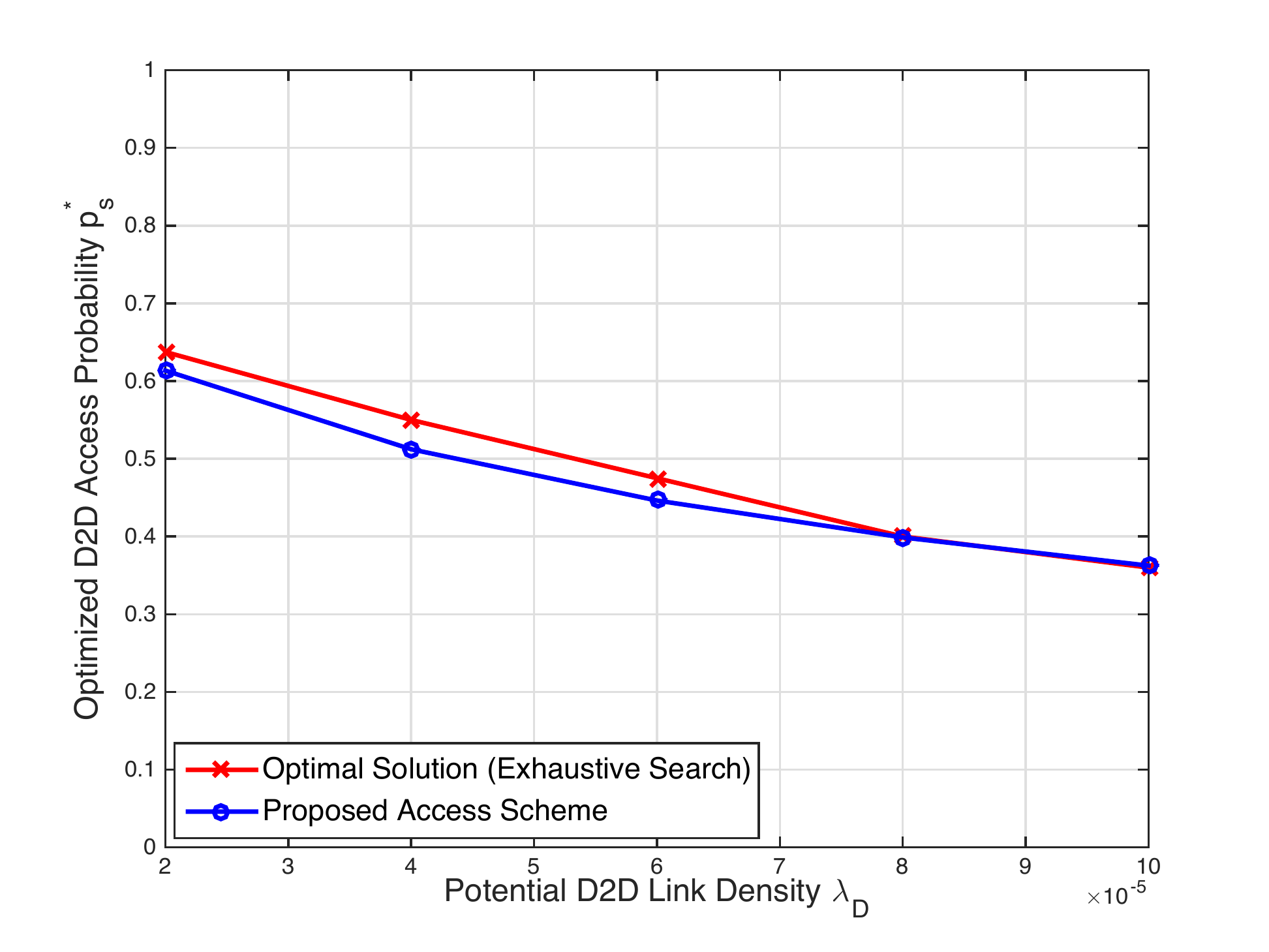}
	\caption{Optimized average access probability $p_s^{\star}$ vs. potential D2D link density $\lambda_D$.}
	\label{optimal_proba}
\end{figure}

\begin{figure}
	\centering
	\includegraphics[scale=0.45]{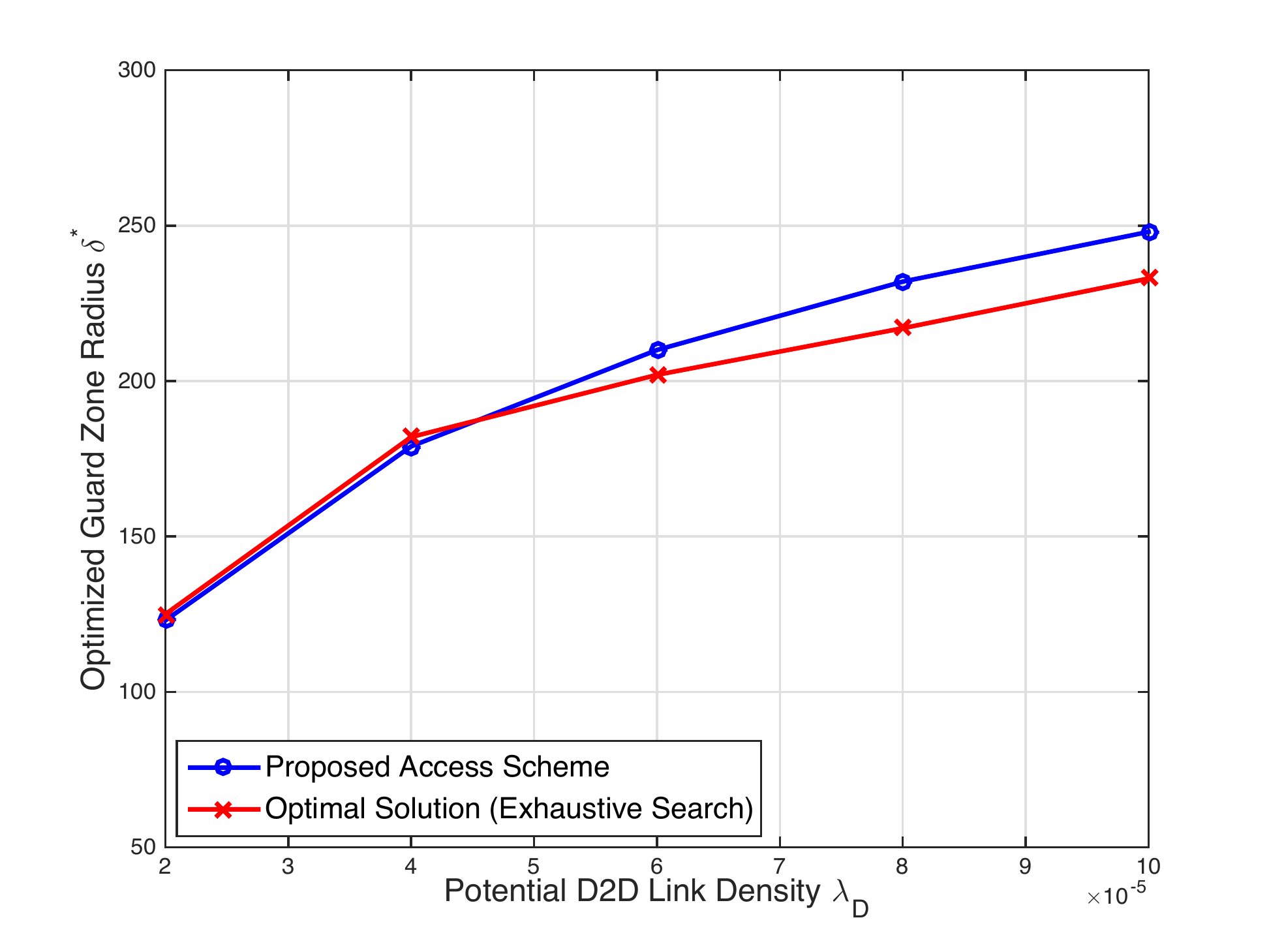}
	\caption{Optimized cellular guard zone radius $\delta^{\star}$  vs. potential D2D link density  $\lambda_D$.}
	\label{delta_vs_lambda}
\end{figure}
In Figs.~\ref{optimal_proba} and \ref{delta_vs_lambda} we compare the optimized D2D access probability $p_s$ and the cellular guard zone radius $\delta$ obtained using the decoupled optimization approach with the optimal solutions obtained by exhaustive search, respectively. The theoretical values of $p_s^{\star}$ are calculated based on Proposition \ref{prop2}, while the $\delta^{\star}$ is obtained from \eqref{optimal_delta_condition}.
We see that our theoretical results of $(p_s^{\star}, \delta^{\star})$ gives relatively close values to the simulated optimal ones. The small gap between the analytical and simulation results is mainly due to the approximation used in Proposition~\ref{theorem_cellular} to calculate the cellular coverage probability. 

\subsection{D2D ASE with Optimized $(p_s, \delta)$ }
\begin{figure}
	\centering
	\includegraphics[scale=0.45]{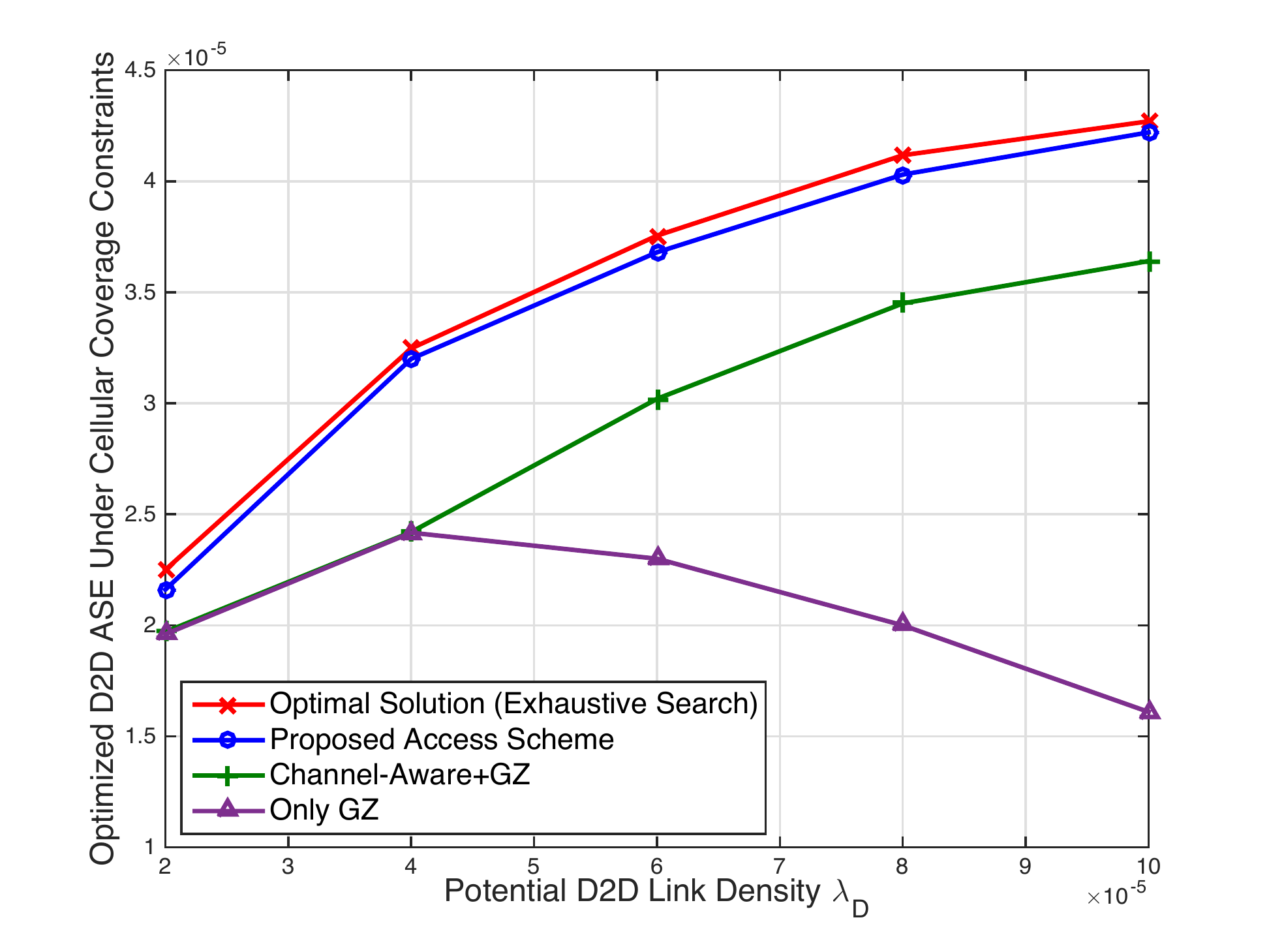}
	\caption{Optimized D2D ASE vs. potential D2D link density $\lambda_D$ with different access control methods.}
	\label{comparison_optimal_ase}
\end{figure}

In Fig. \ref{comparison_optimal_ase}, we evaluate the ASE performance of the D2D tier applying the proposed distributed access control protocol. The results are obtained with $p_s^{\star}$ and $G^{\star}$ as in Proposition \ref{prop2} and guard zone radius $\delta^{\star}$ that satisfies \eqref{optimal_delta_condition}. 
The cellular coverage probability without D2D interference is $p_{\max}^{\text{C}}=0.5552$, implying that the minimum cellular coverage is $p_{\text{cov}}^{\text{C}} \geq (1-\mu)p_{\max}^{C}=0.3886$. 

For comparison, we plot the optimum ASE obtained through exhaustive search, demonstrating that the decoupled approach for optimizing the proposed scheme is very close to the optimal solution. Compared to alternative access control schemes, we observe that our proposed method improves the aggregate throughput and provide evident performance gain. We also see that the SIR-aware access scheme improves the network performance for any range of D2D densities, while the channel-aware method exhibits gains compared to the only GZ scheme starts for densities starting from $\lambda_D=4\times 10^{-5}$. This showcase the importance of taking into account the correlation between the estimated SIR and the real SIR of active D2D links in order to maximize the D2D throughput. 

\begin{table}[htbp]
	\centering
	\caption{Comparison between different D2D access schemes}
	\renewcommand{\arraystretch}{1.5}
	\setlength\tabcolsep{2pt}
	\begin{tabular}{ | c | c | c |c |}
		\hline
		&\footnotesize {D2D Sum Rate} & \footnotesize {Cellular Sum Rate}& \footnotesize {Cellular Coverage}\vspace{-0.1cm}  \\
		&\footnotesize {$R_{\text{D}}~(\times10^{-5})$} &\footnotesize { $R_{\text{C}}~(\times10^{-6})$}& \footnotesize {$p_{\text{cov}}^{\text{C}}$ } \\ \hline
		\footnotesize {Proposed Scheme} & $7.89$ & $1.637$& $0.39064$\\  \cline{1-4}
		\footnotesize {Channel-Aware AC} &$6.55$ &$1.64$   &$0.3966$\\ \cline{1-4}
		\footnotesize {Only Guard Zones} & $5.71$ & $1.656$ &$0.3978$\\ \cline{1-4}
		\footnotesize {No AC} & $7.05$ & $0.455$& $0.094$\\ \hline
	\end{tabular}
	\label{table_data}
\end{table}

\subsection{Average Sum Rate and Cellular Coverage with Optimized $(p_s, \delta)$}
In Table~\ref{table_data} we show the average sum rate per area (bps/Hz/$\text{m}^{2}$) of the D2D tier (denoted by $R_{\text{D}}$) and of the cellular tier (denoted by $R_{\text{C}}$), as well as the cellular coverage probability, achieved with our proposed $p_s^{\star}$ and $\delta^{\star}$ for a given potential D2D link density ($\lambda_D=6\times10^{-5}$). The results are compared with the channel-aware scheme, a scheme implementing only guard zones (step 1 of proposed scheme) and a baseline scheme with no access control. The results evince the performance gains by setting $p_s^{\star}$ and $\delta^{\star}$ according to the decoupled optimization approach. Note that even though the objective of this paper and hence of our optimization problem was to maximize the D2D ASE under cellular coverage constraints, using $p_s^{\star}$ and $\delta^{\star}$ can also improve the average sum rate of D2D network.

\section{Conclusions}
\label{sec:conclusion}
In this work, we proposed a decentralized access control scheme for D2D underlaid cellular networks, which combines SIR-aware link activation with cellular guard zones. Using tools from stochastic geometry, we characterized the impact of the SIR threshold and the exclusion region range on the area spectral efficiency of D2D communications and on the cellular coverage probability. A tractable approach was proposed in order to find the optimal SIR threshold and guard zone radius that maximize the ASE of the D2D tier while guaranteeing sufficient cellular coverage probability. 
The main takeaway of this paper is that very large throughput gains can be achieved in D2D underlaid cellular networks using distributed SIR-aware scheduling in conjunction with cellular exclusion regions. 
Future work could investigate the effect of multiple antennas at the base stations and joint optimization of device association, load balancing and interference avoidance.

\appendix
\appendices
\subsection{Proof of Proposition \ref{propo_laplace} }
\label{appen_laplace}
Due to the asymmetric shape of Voronoi cells, the distribution of the distance from the nearest interfering uplink user to the typical BS is not straightforward. 
From existing results on Poisson-Voronoi tessellations \cite{voronoi, voronoi2}, the area distribution of a Voronoi cell, denoted by $\mathcal{A}$, can be approximated by 
\begin{equation}
f_{\mathcal{A}}(a)=\frac{(3.5\lambda_M)^{3.5}}{\Gamma(3.5)} a^{2.5} \exp(-3.5\lambda_M a).
\end{equation}
If then the typical Voronoi cell is approximated by a circle centered at the typical BS with the same area, the distance from the nearest uplink interferer to the typical cellular receiver (BS) is the radius of the circle. Knowing that $\mathcal{A}=\pi d_{\min}^2$, the distribution of the radius $d_{\min}$ is given by
\begin{equation}
f_{d_{\min}}(r)=2\frac{(3.5\pi \lambda_M)^{3.5}}{\Gamma(3.5)} r^{6} \exp(-3.5\pi \lambda_M r^2).
\label{pdf_dmin}
\end{equation}
From Definition \ref{def_1}, assuming that the distribution of uplink users can be approximated by a homogeneous PPP with density $\lambda_M$, $\mathcal{L}_{I_{cc}}(s)$ can be derived by the Laplace transform of interference coming from PPP-distributed nodes with minimum distance $d_{\min}$ to the typical receiver. Thus we have 
\begin{equation}
\mathcal {L}_{I_{cc}} (s)\approx \mathcal{L}_{I}^{1}(s, \lambda_M, d_{\min}) ,
\end{equation}
where the pdf of $d_{\min}$ is given in \eqref{pdf_dmin}.

\subsection{Proof of Proposition \ref{prop2}}
\label{appen3}
Since $\mathcal{T}(p_{s})$ increases monotonically with $p_{s}$ when $p_{s} \to 0$, and decreases monotonically with $p_{s}$ when $p_{s} \to 1$, and is a continuous function, it is reasonable to consider that the crossing point of these two functions could be approximately the $p_{s}$ that maximizes $\mathcal{T}(p_{s})$. Under this assumption, the optimal access probability $p_{s}^{\star}$ should satisfy
\begin{eqnarray}
p_{s}^{\star} &\simeq&\exp\left[-\xi \beta^{\frac{2}{\alpha}}\left(p_{s}^{\star}\lambda_D+ \kappa \lambda_M \right)\right]\nonumber \\
\Rightarrow  e^{-\xi \beta^{\frac{2}{\alpha}}p_{s}^{\star}\lambda_D}&\simeq&e^{\xi \beta^{\frac{2}{\alpha}}\kappa \lambda_M }p_{s}^{\star} .
\label{optimal_Probability}
\end{eqnarray}
For a general type of equation $p^{ax+b}=cx+d$, where $x$ is the variable and $a$, $b$, $c$, $d$, $p$ are constant, when $p>0$ and $a,c\neq 0$, the solution by using Lambert $W$ function is
\begin{equation}
x=-\frac{\mathcal{W}\left(-\frac{a\ln p}{c} p^{b-\frac{ad}{c}}\right)}{a\ln p}-\frac{d}{c}.
\label{lambert}
\end{equation}
By solving \eqref{optimal_Probability} with the help of Lambert W funtion we have
\begin{equation}
p_{s}^{\star}\simeq\frac{\mathcal{W}\left(\lambda_D \xi \beta^{\frac{2}{\alpha}}  e^{-\kappa \lambda_M \xi \beta^{\frac{2}{\alpha}}}\right)}{\lambda_D \xi \beta^{\frac{2}{\alpha}}}.
\label{conditional_Ps}
\end{equation}
Knowing that $p_{s}^{\star}$ should not exceed one, we have
\begin{equation}
p_{s}^{\star}\simeq\min\left\{\frac{\mathcal{W}\left(\lambda_D \xi \beta^{\frac{2}{\alpha}}  e^{-\kappa \lambda_M \xi \beta^{\frac{2}{\alpha}}}\right)}{\lambda_D \xi \beta^{\frac{2}{\alpha}}},1\right\}.
\end{equation} 
Substituting it into $\eqref{Ps}$, we have that the approximately optimal SIR threshold is given as
\begin{equation}
G^{\star}\simeq\left[\frac{-\ln p_{s}^{\star} }{\xi (\lambda_D+\kappa \lambda_M)}\right]^{\frac{\alpha}{2}}.
\end{equation}


\end{document}